\documentclass[reqno,centertags]{amsart}

\usepackage[sort&compress]{natbib}
\usepackage{amsmath}
\usepackage{psfrag}
\usepackage[dvips]{graphicx}
\usepackage{latexsym}

\newcommand{\Frac}[2]{{{\displaystyle#1}\over{\displaystyle#2}}}%
\newcommand{\pder}[2]{{{\displaystyle\partial#1}\over{\displaystyle\partial#2}}}%
\newcommand{\der}[2]{\Frac{\mathrm{d}#1}{\mathrm{d}#2}}%
\newcommand{\dertwo}[2]{\Frac{\mathrm{d}^{2}#1}{\mathrm{d}{#2}^{2}}}%
\newcommand{\matder}[1]{\left(v_{x}\pder{}{x} + v_{y}\pder{}{y}\right)}
\newcommand{\Int}[4]{\int_{#1}^{#2}#3\,\mathrm{d}#4}%
\newcommand{\textfrac}[2]{\ensuremath{#1/#2}}%
\newcommand{\textfracp}[2]{\ensuremath{(#1/#2)}}%
\newcommand{\textder}[2]{\textfrac{\mathrm{d}#1}{\mathrm{d}#2}}%
\newcommand{\textpder}[2]{\textfrac{\partial#1}{\partial#2}}
\newcommand{\ie}{i.e.}%
\newcommand{\fig}[1]{figure~#1}%
\newcommand{\mytable}[1]{(Table~{#1})}%
\newcommand{\BlackBox}{\protect\rule{0.6em}{0.6em}}
\newcommand{\OrderOf}[1]{\textit{O}(#1)}
\newcommand{\Vector}[1]{\underline{#1}}
\newcommand{\Tensor}[1]{\underline{\underline{#1}}}
\newcommand{\mysqrtp}[1]{{\left(#1\right)}^{1/2}}
\newcommand{\mysqrt}[1]{{#1}^{1/2}}
\newcommand{\eqnannounce}[1]{%
  \par\pagebreak[2]\noindent\begingroup#1\endgroup}
\newcommand{\Modulus}[1]{\ensuremath{\lvert#1\rvert}}
\newcommand{\inner}[1]{#1^{\prime}}

\newcommand{\Inner}[1]{#1''}
\newcommand{\xibar}{\ensuremath\overline{\xi}}
\newcommand{\airyai}{\ensuremath\mathrm{Ai}}
\newcommand{\airybi}{\ensuremath\mathrm{Bi}}

\providecommand{\BlackBox}{\protect\rule{0.6em}{0.6em}}
\newcommand{\mypsfrag}[2]{\psfrag{#1}[1][1]{\shortstack{#2}}}
\newcommand{\mypsfragr}[2]{\psfrag{#1}[r][1][1][270]{\shortstack{#2}}}

\newcommand{\linezero}{\protect\rule[0.4ex]{3.5em}{0.3ex}}

\newcommand{\lone}{\protect\rule[0.4ex]{0.15cm}{0.3ex}\kern.15cm}
\newcommand{\lineone}{\lone\lone\lone\lone\lone\kern-.15cm}

\newcommand{\ltwo}{\protect\rule[0.4ex]{0.225cm}{0.3ex}\kern.075cm}
\newcommand{\linetwo}{\ltwo\ltwo\ltwo\ltwo\ltwo\kern-.075cm}

\newcommand{\lthree}{\protect\rule[0.4ex]{0.075cm}{0.3ex}\kern.225cm}
\newcommand{\linethree}{\lthree\lthree\lthree\lthree\lthree\lthree\kern-.225cm}

\newcommand{\dotted}{\kern.07cm\protect\rule[0.4ex]{0.07cm}{0.3ex}}
\newcommand{\dotone}{\dotted\dotted\dotted}
\newcommand{\linedotted}{\kern-.07cm\dotone\dotone\dotone}

\newcommand{\dotdash}{\kern.15em\protect\rule[0.4ex]{0.15em}{0.3ex}\kern.15em%
  \protect\rule[0.4ex]{0.6em}{0.3ex}}
\newcommand{\linedotdash}{\dotdash\dotdash\dotdash\dotdash}

\bibpunct{[}{]}{,}{n}{}{,}
\begin{document}

\title[A frictional Cosserat model for granular materials]{%
  A frictional Cosserat model for the flow of granular materials through
  a vertical  channel}
\author[L. S. Mohan]{\textbf{L. Srinivasa Mohan}}
\author[P. R. Nott]{\textbf{Prabhu R. Nott}}
\author[K. K. Rao]{\textbf{K. Kesava  Rao}}

\date{}

\begin{abstract}
  A rigid-plastic Cosserat model has been used to study dense, fully
  developed flow of granular materials through a vertical channel.
  Frictional models based on the classical continuum do not predict
  the occurrence of shear layers, at variance with experimental
  observations. This feature has been attributed to the absence of a
  material length scale in their constitutive equations.  The present
  model incorporates such a material length scale by treating the
  granular material as a Cosserat continuum. Thus localised couple
  stresses exist and the stress tensor is asymmetric. The velocity
  profiles predicted by the model are in close agreement with
  available experimental data.  The predicted dependence of the shear
  layer thickness on the width of the channel is in reasonable
  agreement with data. In the limit of small $\epsilon$ (ratio of the
  particle diameter to the half-width of the channel), the model
  predicts that the shear layer thickness scaled by the particle
  diameter grows as $\epsilon^{\textfrac{-1}{3}}$.
\end{abstract}

\maketitle

\section{Introduction}
\label{sec:intro}

In many terrestrial flows of granular materials, gravity consolidates
the medium to a state where sustained frictional contact between the
particles is the dominant mode of momentum transfer.  In this regime
of high solids fraction and low deformation rate, models based on
concepts in metal plasticity and soil mechanics have been
traditionally used to describe the flow~\citep{jackson83}.  While many
gross features of granular flows can be predicted using these models,
one aspect they fail to capture is the thickness of shear layers;
often when granular materials are sheared, large portions of the
material do not suffer sustained deformation. In the experiments of
\citet{roscoe70, neddermanandlaohakul80, gudehusandtejchman91}, the
velocity gradients are confined to layers approximately 5--40 particle
diameters in thickness.  Moreover, the thickness of the shear layers
is influenced by the nature of the boundaries; when the flowing medium
is confined by smooth walls, it is found that the thickness of the
shear layers is less than that in the case of rough walls
\citep{neddermanandlaohakul80, tejchmanandgudehus93}.

Conventional models of plasticity do not predict shear
layers~\citep{tejchmanandwu93, mohanetal97}. The failure of the
frictional models to predict the thickness of the shear layers
accurately has been attributed to the absence of a material length
scale in their constitutive equations~\citep[cited
in~\citealp{muhlhausandvardoulakis87}]{muhlhaus86}.  To overcome this
deficiency of the classical models, the particle size must be
incorporated in the constitutive equations. In the absence of a
comprehensive micro-mechanical model to describe friction, a continuum
theory that includes a material length scale in the constitutive
equations can be constructed by modelling the granular material as a
Cosserat continuum~\citep{muhlhaus86}.  We shall argue later that the
frictional nature of particle interactions provides sufficient grounds
for using this approach. We note here that models based on kinetic
theory~\citep[see, for example,][]{lunetal84}, involve the particle
diameter in the constitutive relations.  However, these models are
expected to hold only for rapid flows, where particle interactions may
be approximated by instantaneous collisions.

In this paper, we explore the use of a Cosserat plasticity model to
describe steady, fully developed, plane flow of a granular material in
a vertical channel under the action of gravity.  The predictions of
the model will be compared with data reported in the literature. While
Cosserat plasticity models have been applied to problems in granular
flow in the past
\citep{muhlhausandvardoulakis87,muhlhaus89,tejchmanandwu93,%
  tejchmanandgudehus93,tejchmanandwu94}, these studies address
unsteady flows and are posed in terms of strain increments; with this
formulation, \citet{tejchmanandgudehus93} found it difficult to
integrate the equations numerically for large times.  To the best of
our knowledge, the present work represents the first attempt to
examine steady flow in this context.  We indicate below how the model
is developed, and then apply it to channel flow.

\section{The Cosserat model}
\label{sec:cosserat}

\begin{figure}
\label{fig:1}
  \begin{center}
    \includegraphics[bb=200 400 389 734]{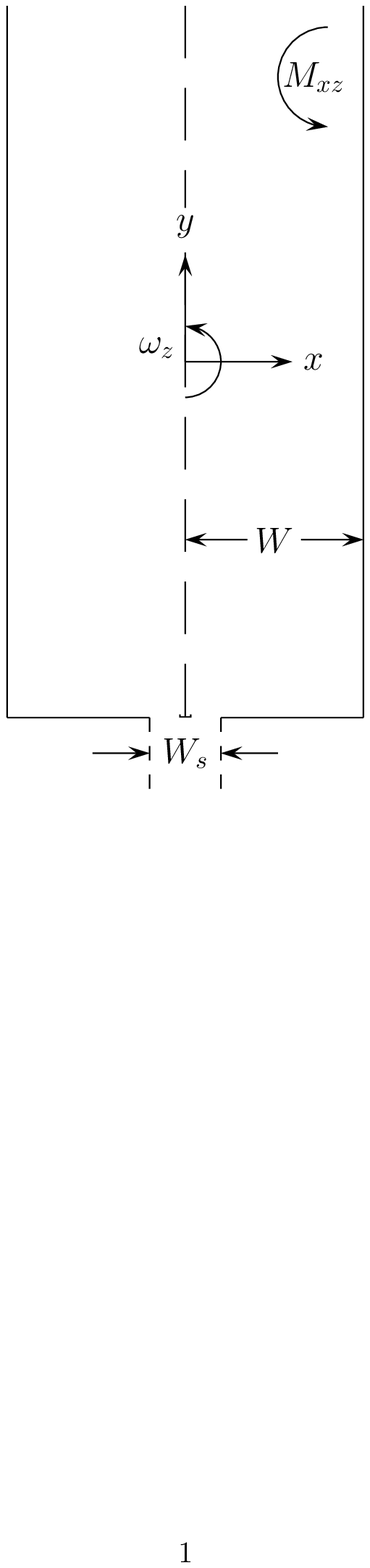}
    \caption{Elevation of the channel. The angular velocity
      $\omega_{z}$ is positive for an anti-clockwise rotation about
      the $z$-axis. The couple stress $M_{xz}$ exerted on the plane
      $x=\mathrm{constant}$ by the material to the left of this plane
      is positive when the couple is directed as shown.}
  \end{center}
\end{figure}

The field variables of the classical continuum are the density $\rho$,
the linear velocity $\Vector{v}$, and the stress tensor
$\Tensor{\sigma}$.  A Cosserat
continuum~(\citep[p.~223]{jaunzemis67}; \citep{cowin74}) involves
two additional field variables, namely, the angular velocity
$\Vector{\omega}$, and the couple stress tensor $\Tensor{M}$.
Considering a Cartesian coordinate system (\fig{1}), $M_{xz}$
represents the couple per unit area exerted about the $z$-axis on a
plane~$x = \mathrm{constant}$, by the material to the left of this
plane.  A positive value of $M_{xz}$ is taken to impose an
anti-clockwise rotation on this plane~(\fig{1}).  For a Cosserat
continuum, the mass and linear momentum balances must be supplemented
by the angular momentum balance, which relates $\Vector{\omega}$,
$\Tensor{M}$, and $\Tensor{\sigma}$. For steady, fully developed flow,
spatial gradients of $\Tensor{M}$ cause $\Tensor{\sigma}$ to be
asymmetric. This is in contrast to the classical continuum, which
assumes implicitly that there are no couple stresses, body couples,
and intrinsic angular momentum; hence the angular momentum balance can
be satisfied identically by requiring $\Tensor{\sigma}$ to be
symmetric.

There is enough analytical evidence to motivate the use of a Cosserat
model in the present problem. \citet{dahler59} used a statistical
mechanical approach to develop expressions for the stresses in a fluid
composed of diatomic molecules. For molecules interacting via central
forces, which are directed along the lines joining the centers of mass
of the molecules, $\Tensor{\sigma}$ is found to be symmetric. However,
his model suggests that non-central forces may cause $\Tensor{\sigma}$
to be asymmetric. \citet{campbell93a} simulated the shearing of
circular discs between parallel plates, assuming that the collisions
between discs, and between a disc and the wall, were instantaneous. In
the latter case, wall roughness was incorporated by imposing (after
collision) either (i)~a zero relative velocity between the surface of
the disc and the wall, or~(ii)~a zero relative velocity between the
center of the disc and the wall. In both cases $\Tensor{\sigma}$ was
asymmetric near the wall, and there were non-zero couple stresses.

\citet{jenkins89} constructed a micro-mechanical model for an assembly
of identical spheres. They found that asymmetric stresses resulted
when the distribution of contact normals was anisotropic; however,
they secured the symmetry of the stress tensor by suitably enforcing
the rotation of particles.

Dry friction, the dominant mode of momentum transfer in high-density
flows, introduces non-central forces in an inherently complex fashion.
Hence, we expect that a micro-mechanical model for dry friction would
result in a continuum with asymmetric stresses; such materials can be
modelled as Cosserat continua. A satisfactory micro-mechanical model
is not yet available, and it is hoped that this issue will be
addressed by future investigators.

\subsection{Equations of motion}
\label{sec:eqnofmotion}

It is instructive to write the equations for the case of steady plane
flow, and later simplify them for the case of fully developed flow.
For flow parallel to the $xy$ plane~(\fig{1}), the velocity field has
the following form:
\begin{equation}
  \label{e-1}
  v_{x} = v_{x} (x, y); \; v_{y} = v_{y} (x, y); \; v_{z} = 0; \;
  \omega_{x} = \omega_{y} = 0; \; \omega_{z} = \omega_{z} (x, y),
\end{equation}
where $v_{x}$ and $\omega_{x}$ are the $x$ components of the linear
and angular velocity, respectively. A positive value of $\omega_{z}$
is associated with an anti-clockwise rotation about the $z$-axis.

The balances for mass and linear momentum are
\begin{align}
  \label{e-5}
  \pder{}{x} (\nu v_{x}) + \pder{}{y} (\nu v_{y}) &= 0,\\
  \label{e-6}
  \pder{\sigma_{xx}}{x} + \pder{\sigma_{yx}}{y} +
  \rho_{p}\nu\matder{}\:v_{x} &= 0,\\
  \label{e-7}
  \pder{\sigma_{xy}}{x} + \pder{\sigma_{yy}}{y} +
  \rho_{p}\nu\matder{}\:v_{y} &= - \rho_{p}
  \nu g,
\end{align}
where $\nu$ is the solids fraction or the volume fraction of solids,
$\sigma_{ij}$'s are the components of the Cauchy stress tensor,
defined in the compressive sense, $\rho_{p}$ is the intrinsic
density of the particles, assumed constant, and $g$ is the
acceleration due to gravity.

Following~\citet[p.~233]{jaunzemis67}, the $z$ component of the
angular momentum balance is
\begin{equation}
  \label{e-8}
  \pder{M_{xz}}{x} + \pder{M_{yz}}{y} - \rho_{p} \nu
  \zeta_{z} + \sigma_{xy} - \sigma_{yx} + \rho_{p}\nu\matder{}\:\eta_{z}
  = 0,
\end{equation}
where $M_{iz}$'s are the couple stresses, $\eta_{z}$ is the $z$
component of the intrinsic angular momentum (per unit volume), and
$\zeta_{z}$ is the $z$ component of the body couple acting on the
material.

To close the above set of equations, constitutive relations for the
$\sigma_{ij}$ and $M_{iz}$ are required.

\subsection{Constitutive equations}
\label{sec:consteqn}

\citet{muhlhausandvardoulakis87} and~\citet{tejchmanandwu93} have
developed Cosserat plasticity models for studying the development of
shear bands in granular flow. In their models, the yield condition and
the flow rule were modified to account for the influence of the couple
stress and to provide a relation for the angular velocity. We have
adapted their model to the present problem. The constitutive equations
comprise of a yield condition and a flow rule, which are elaborated
below.

\subsubsection{Yield condition}

Following~\citet[cited in~\citealp{lippmann95}]{besdo74},
\citet{deBorst93}, and~\citet{tejchmanandwu93}, we use a yield
condition of the form
\begin{equation}
  \label{e-9}
  F \equiv \tau  -  Y  = 0,
\end{equation}
where
\begin{equation}
  \label{e-10}
  \tau \equiv \mysqrt{\left(a_{1} \sigma_{ij}' \sigma_{ij}' + a_{2}
      \sigma_{ij}' \sigma_{ji}' + \Frac{1}{(L d_{p})^{2}} M_{ij}
      M_{ij} \right)},
\end{equation}
$\sigma_{ij}' = \sigma_{ij} - \textfracp{1}{3}\,\sigma_{kk}\,
\delta_{ij}$ is the deviatoric stress, $\delta_{ij}$ is the
Kronecker delta, $a_{1}$, $a_{2}$, and~$L$ are material constants,
and $d_{p}$ is the particle diameter. Here $L\,d_{p}$ is a
characteristic material length scale; the value of $L$ will be
chosen later. \citet{deBorst93} assumes that the yield limit $Y$
depends on the mean stress $\sigma\equiv\textfrac{\sigma_{kk}}{3}$,
and a hardening parameter $h$. Here we identify $h$ with the solids
fraction $\nu$.

\citet[p.~112]{schofieldandwroth68} and~\citet{jackson83} discuss
the use of a yield condition of the form $F(\Tensor{\sigma}, \nu) =
0$ in the classical frictional models. The yield
condition~\eqref{e-9} with $Y = Y(\sigma, \nu)$ represents an
attempt to include couple stresses within this framework. Only two
of the three parameters $a_{1}$, $a_{2}$ and $L$ in~\eqref{e-10} are
independent, because the third parameter may be absorbed in the
definition of $Y(\sigma,\nu)$~(see~\eqref{e-9}).  Following
\citet{muhlhausandvardoulakis87} we set $a_{1} + a_{2} =
\textfrac{1}{2}$, without loss of generality.

\citet{tejchmanandwu93} use $\mathrm{A}\equiv\textfrac{a_{2}}{a_{1}}
= \textfrac{1}{3}$ and $L = 1$. Here we retain their choice of
$\mathrm{A}$ and and treat $L$ as an adjustable parameter, whose
value is chosen as described later.  \citet{deBorst93} found that
changes in the values of $\mathrm{A}$ and $L$ affected the post-peak
behaviour of a sample which was sheared between parallel plates.
Unfortunately, neither experiments nor satisfactory micro-mechanical
models are available to guide the choice of $\mathrm{A}$.

Following~\citet{prakashandrao88}, we assume the following form for
the yield limit $Y$
\begin{equation}
  \label{e-11}
  Y = \sigma_{c}(\nu) \sin \phi
  \left(
    n \alpha - (n - 1) \alpha^{\textfracp{n}{(n - 1)}}
  \right); \quad \alpha\equiv\Frac{\sigma}{\sigma_{c}(\nu)}.
\end{equation}
Here $\sigma_{c}(\nu)$ is the mean stress at a critical state, $\phi$
is a material constant called the angle of internal friction, and $n$
is a material constant. The significance of a critical state will be
explained shortly.  The dependence of $\sigma_{c}$ on the solids
fraction $\nu$ is taken to be~\citep{johnsonandjackson87}
\begin{equation}
  \label{eq:24}
  \sigma_{c}(\nu) =
  \begin{cases}
    0 &\nu < \nu_{\min},\\
    \Lambda\Frac{(\nu - \nu_{\min})^{p}}{(\nu_{\max} -
      \nu)^{q}} &\nu_{\min} \leq \nu \leq \nu_{\max}.
  \end{cases}
\end{equation}
Here $\Lambda$, $\nu_{\min}$, $\nu_{\max}$, $p$ and $q$ are material
constants.  Note that $\sigma_{c}(\nu)$ has been chosen to be zero
below $\nu_{\min}$, the solids fraction at loose random packing, and
to diverge as $\nu$ approaches $\nu_{\max}$, the solids fraction at
dense random packing.

\subsubsection{Flow rule}

\citet{tejchmanandwu93} have used incremental elasto-plastic
constitutive equations, which they attribute to~\citet{muhlhaus89}.
Elastic effects are ignored in the present work to simplify the
analysis. Because we are interested in sustained flow, the plastic
strain increments used by~\citet{tejchmanandwu93} are replaced by
suitable velocity gradients~\citep{muhlhaus89}. In Cartesian tensor
notation, the flow rule is given by
\begin{equation}
  \label{eq:22}
  E_{ij} \equiv \pder{v_{i}}{x_{j}} + \varepsilon_{ijk}\omega_{k} =
  \dot{\lambda} \pder{G}{\sigma_{ji}}; \quad H_{ij} \equiv
  \pder{\omega_{i}}{x_{j}} = \dot{\lambda}\pder{G}{M_{ji}},
\end{equation}
where $G(\Tensor{\sigma}, \Tensor{M}, \nu)$ is the plastic potential,
$\varepsilon_{ijk}$ is the alternating tensor, and $\dot{\lambda}$ is
a scalar factor. We note here that $E_{ij}$ is the sum of the rate of
deformation tensor $D_{ij}$ and an objective antisymmetric tensor
representing the difference between the spin tensor and the particle
spin $\varepsilon_{ijk}\omega_{k}$.  $E_{ij}$ and $H_{ij}$ are
conjugate to the stress $\sigma_{ji}$ and the couple stress $M_{ji}$,
respectively, in the sense that the rate of working per unit volume by
the contact forces and couples is given by $-(\sigma_{ji}E_{ij} +
M_{ji}H_{ij})$~\citep{muhlhaus89}. In a classical continuum,
$\Tensor{M}$ vanishes and $\Tensor{\sigma}$ is symmetric; hence the
above expression reduces to $-\sigma_{ji}D_{ij}$, where $D_{ij} =
\textfracp{1}{2}\:(\textpder{v_{i}}{x_{j}} + \textpder{v_{j}}{x_{i}})$
denotes a component of the rate of deformation tensor.

Lacking detailed information on the plastic potential $G$, we adopt
the most commonly used closure, namely the associated flow rule
\citep[p.~43]{schofieldandwroth68}:
\begin{equation}
  \label{e-17}
  G \equiv F = \tau - Y.
\end{equation}
This form of the flow rule accounts for density changes accompanying
deformation.

\section{Application to channel flow}
\label{sec:problem}

For the case of steady, fully developed, plane flow, the velocity
field is given by
\begin{equation}
  v_{y} = v_{y}(x); \quad \omega\equiv\omega_{z}(x),\label{e-35}
\end{equation}
and the other velocity components vanish. Hence the mass
balance~\eqref{e-5} is identically satisfied and the balances of
linear and angular momentum~\eqref{e-6}--\eqref{e-8} reduce to
\begin{gather}
  \der{\sigma_{xx}}{x}  =  0;\quad\der{\sigma_{xy}}{x}  =  -
  \rho_{p} \nu g,\label{e-20}\\
  \der{m}{x} + \sigma_{xy} - \sigma_{yx} = 0,
  \label{e-21}
\end{gather}
where $m \equiv M_{xz}$. It is assumed that the yield condition is
satisfied at every point in the channel, so that the factor
$\dot{\lambda}$ in~\eqref{eq:22} is always non-zero. In
writing~\eqref{e-21}, it is assumed that there are no body couples.

\subsection{The stress field}
\label{sec:Some-cons-assum}

For fully developed flow, it will now be shown that all the normal
stresses are equal. Equation~\eqref{eq:22} implies that
\begin{equation}
  \label{e-27}
  E_{xx} = \pder{v_{x}}{x} = 0 = \Frac{\dot{\lambda}}{6\tau}
  (2 \sigma_{xx}' - \sigma_{yy}' - \sigma_{zz}') -
  \Frac{\dot{\lambda}}{3}\pder{Y}{\sigma}.
\end{equation}
Writing the corresponding equations for $E_{yy}$ and $E_{zz}$ and
summing, we get
\begin{equation}
  \label{eq:23}
  \pder{Y}{\sigma} = 0,
\end{equation}
or using~(\ref{e-11})
\begin{equation}
  \label{eq:31}
  \sigma = \sigma_{c}(\nu); \quad Y = \sigma_{c} \sin \phi.
\end{equation}
Thus the material is at a critical state or a state of isochoric
deformation, because $E_{ii} = \nabla \cdot \Vector{v} = 0$.
Comparison of \eqref{e-11} and \eqref{eq:31} shows that the value of
$n$ is not relevant. It also follows from~\eqref{e-27}, \eqref{eq:31}
and~\eqref{e-20} that
\begin{equation}
  \label{eq:33}
  \sigma_{xx} = \sigma_{yy} = \sigma_{zz} = \sigma_{c}(\nu) =
  \mathrm{constant}.
\end{equation}
Hence the solids fraction does not vary across the width of the
channel.

Using~\eqref{eq:31} and \eqref{eq:33}, the yield condition~\eqref{e-9}
reduces to
\begin{equation}
  \label{eq:34}
  \mysqrtp{a_{1}(\sigma_{xy}^{2} + \sigma_{yx}^{2}) + 2 a_{2}
  \sigma_{xy}\sigma_{yx} + \Frac{m^{2}}{{(Ld_{p})}^{2}}} -
  \sigma_{c}(\nu)\sin\phi = 0.
\end{equation}

\subsection{Velocity field}
\label{sec:Const-relat}

The non-trivial equations of the flow rule~\eqref{eq:22} are
\begin{eqnarray}
  E_{xy} & = & \omega = \Frac{\dot{\lambda}}{\tau}
  \left(a_{1} \sigma_{yx} + a_{2} \sigma_{xy}
  \right),\label{e-32}\\
  E_{yx} & = & \der{v_{y}}{x} - \omega =
  \Frac{\dot{\lambda}}{\tau} \left(a_{1} \sigma_{xy} + a_{2}
    \sigma_{yx}
  \right),\label{e-33}\\
  H_{zx} & = & \der{\omega}{x} =
  \Frac{\dot{\lambda}}{\tau} \Frac{m}{(L d_{p})^{2}}.
  \label{e-34}
\end{eqnarray}

On eliminating $\dot{\lambda}$ we get
\begin{eqnarray}
  \der{v_{y}}{x}& = & \Frac{(\mathrm{A} + 1)(\sigma_{xy} +
    \sigma_{yx})\,\omega}{\sigma_{yx} + \mathrm{A}
    \sigma_{xy}},\label{e-46}\\
  \der{\omega}{x} & = &
  \Frac{2(\mathrm{A} + 1)\,m\,\omega}{(L d_{p})^{2} (\sigma_{yx} + \mathrm{A}
    \sigma_{xy})}. \label{e-45}
\end{eqnarray}

\subsection{Boundary conditions}
\label{sec:Boundary-conditions-2}

Considering symmetric solutions, we have
\begin{equation}
  \sigma_{xy}(0)  =  0;\qquad \omega(0) =  0.\label{e-23}
\end{equation}
The angular velocity $\omega$ must vanish at the centerline of the
channel, because a non-zero value implies a preferred direction of
rotation.

Equations~\eqref{e-32} and~\eqref{e-23} imply
\begin{equation}
  \label{eq:11}
  \sigma_{yx}(0) = 0,
\end{equation}
provided $m(0)$ is bounded, \ie, $\nu<\nu_{\max}$ (see \eqref{e-10},
\eqref{eq:24}, and \eqref{eq:34}).  Because $\sigma_{xy}$ and
$\sigma_{yx}$ both vanish at the centerline, the yield
condition~\eqref{eq:34} implies that the couple stress at the
centerline is
\begin{equation}
  \label{eq:12}
  m(x = 0) = \pm L d_{p}\sigma_{c}
  \sin \phi.
\end{equation}
While both roots in~\eqref{eq:12} are mathematical solutions, only the
negative root yields a physically reasonable solution.  The
justification for choosing this root, and the reason for discarding
the other are discussed in Appendix~A.

At the right hand wall $x = W$ we use the usual friction boundary
condition~(\citep{brennenandpearce78}; \citep[p.~40]{nedderman92})
\begin{equation}
  \label{eq:5}
  -\Frac{\sigma_{xy}}{\sigma_{xx}} = \tan \delta \quad \text{at} \quad
  x = W,
\end{equation}
where $\delta$ is a constant called the angle of wall friction.
Using~\eqref{e-20} and~\eqref{eq:33}, \eqref{eq:5} reduces to
\begin{equation}
  \label{eq:35}
  \Frac{\rho_{p}\nu g W}{\sigma_{c}(\nu)} = \tan\delta \quad \text{at} \quad
  x = W,
\end{equation}
which determines the value of $\nu$ for specified values of $W$ and
$\delta$.

Following~\citet{tejchmanandgudehus93}, we assume that
\begin{equation}
  \label{e-63}
  v_{y} = -Kd_{p}\,\omega \quad \text{at} \quad x = W,
\end{equation}
where $K$ is a dimensionless constant which reflects the roughness of
the wall. To get a feel for this condition, consider a single
spherical particle sliding or rolling down a vertical wall.  Let
$v_{y}'$ and $\omega{'}$ represent the linear velocity of the center
of the particle and its angular velocity about an axis through its
center, respectively. If the particle slides without rolling,
$\omega{'} = 0$, but $v_{y}'$ is arbitrary.  Conversely, if it rolls
without slipping
$\Modulus{v_{y}'}=\textfracp{d_{p}}{2}\:\Modulus{\omega{'}}$.  For the
boundary condition~\eqref{e-63}, these limits correspond to $K
\rightarrow \infty$ and $K \rightarrow \textfrac{1}{2}$, respectively.
Reverting to the continuum, we expect that $K$ will decrease as the
wall roughness increases.

One more boundary condition is needed to permit determination of all
the integration constants. Here we set
\begin{equation}
  \label{eq:26}
  \omega(x = W) = \omega_{w},
\end{equation}
where $\omega_{w}$ is a constant whose value is determined by
adjusting either the mass flow rate or the centerline velocity to
match the measured value. In experiments, the mass flow rate may be
varied within limits by varying the width $2 W_{s}$ of the exit slot
at the bottom of the channel. Because we are considering fully developed
flow, $W_{s}$ does not occur explicitly in the governing equations,
but its influence is incorporated by changing $\omega_{w}$.

\section{Solution procedure}
\label{sec:solution}

Introducing the dimensionless variables
\begin{align}
  \xi & = \Frac{x}{W}; & \epsilon
  &=\Frac{d_{p}}{W}; & u &= -
  \Frac{v_{y}}{\mysqrtp{g W}}; \notag\\
  \overline{\omega} &= \omega\left(\Frac{W}{g}\right)^{\textfrac{1}{2}};
  &\overline{\sigma}_{ij} &= \Frac{\sigma_{ij}}{\rho_{p} g W}; &
  \overline{m} &= \Frac{m}{\rho_{p} g W d_{p}},\notag
\end{align}
the balance equations~\eqref{e-20} and~\eqref{e-21} may be rewritten as
\begin{align}
  \der{\overline{\sigma}_{xx}}{\xi}  &=  0,\label{e-54}\\
  \der{\overline{\sigma}_{xy}}{\xi}  &=  - \nu,\label{e-55}\\
  \epsilon\der{\overline{m}}{\xi} + \overline{\sigma}_{xy} -
  \overline{\sigma}_{yx} &=  0,\label{e-61}
\end{align}
where $\nu$ is the constant solids fraction across the channel, and
\begin{equation}
  \label{eq:2}
  \epsilon\equiv\Frac{d_{p}}{W}.
\end{equation}

The dimensionless form of the yield condition~\eqref{eq:34} is
\begin{equation}
  \label{eq:8}
  a_{1}(\overline{\sigma}_{xy}^{2} + \overline{\sigma}_{yx}^{2}) + 2
  a_{2} \overline{\sigma}_{xy} \overline{\sigma}_{yx} +
  \Frac{\overline{m}^{2}}{L^{2}} =
  (\overline{\sigma}_{c}\sin\phi)^{2}.
\end{equation}
Here $\overline{\sigma}_{c}(\nu) = \textfrac{\sigma_{c}}{(\rho_{p} g
  W)}$.  The flow rule~\eqref{e-46} and~\eqref{e-45} is given by
\begin{align}
  \der{u}{\xi} &= - \Frac{(1 + \mathrm{A})(\overline{\sigma}_{xy} +
    \overline{\sigma}_{yx})\,\overline{\omega}}
  {\overline{\sigma}_{yx} + \mathrm{A}
    \overline{\sigma}_{xy}},\label{e-59}\\
  \der{\overline{\omega}}{\xi} &= \Frac{2(1 + \mathrm{A})\,
    \overline{m}\,\overline{\omega}}{\epsilon L^{2}
    (\overline{\sigma}_{yx} +
    \mathrm{A}\overline{\sigma}_{xy})}.\label{e-60}
\end{align}
The boundary conditions are:
\eqnannounce{at the centerline ($\xi = 0$)}
\begin{equation}
  \overline{\sigma}_{xy} = 0; \quad \overline{\omega} = 0.\label{eq:6}
\end{equation}
\eqnannounce{at the wall ($\xi = 1$)}
\begin{equation}
  \Frac{\nu}{\overline{\sigma}_{c}(\nu)} = \tan \delta; \quad u = \epsilon K
  \overline{\omega}.\label{eq:7}
\end{equation}

Equations~\eqref{eq:11} and~\eqref{eq:12} may be written as
\begin{equation}
  \label{eq:54}
  \overline{\sigma}_{yx} = 0, \quad \overline{m} =
  -NL\nu \quad \text{at}
  \quad \xi = 0,
\end{equation}
where $N\equiv\textfrac{\sin\phi}{\tan\delta}$, and~\eqref{eq:7} has been
used to simplify the second of equations~\eqref{eq:54}.

\subsection{Method of solution}
\label{sec:Method-solution}

For the special case $\mathrm{A} = -1$, an analytical solution may be
obtained as discussed in Appendix~A. This shows that
$\overline{m}=\mathrm{constant}$,
$\overline{\sigma}_{yx}=\overline{\sigma}_{xy}$, and
$\overline{\omega}$ and $u - u(0)$ display a power law dependence on
$\xi$. (The case $\mathrm{A}<-1$ is discussed in Appendix~A.)

We now discuss the case $\mathrm{A}>-1$. Inspection
of~\eqref{e-54}--\eqref{eq:54} shows that the stress field is
uncoupled from the velocity field. Hence we first integrate the
equations~\eqref{e-54},~\eqref{e-55} and~\eqref{eq:6}.
Equations~\eqref{e-54} and~\eqref{e-55} may be solved along with the
first of boundary conditions~\eqref{eq:6} to get
\begin{equation}
  \label{eq:19}
  \overline{\sigma}_{xx} = \mathrm{constant}; \quad
  \overline{\sigma}_{xy} = -\nu\xi.
\end{equation}
The yield condition~\eqref{eq:8} may be solved for
$\overline{\sigma}_{yx}$ to get
\begin{equation}
  \label{eq:56}
  \overline{\sigma}_{yx} = \mathrm{A}\nu\xi \mp
    \mysqrtp{(\mathrm{A}^{2} - 1)(\nu\xi)^{2} +
      2(\mathrm{A} + 1)\left(N^{2}\nu^{2} -
    \Frac{\overline{m}^{2}}{L^{2}}\right)}.
\end{equation}
In our calculations, only the root with the negative sign before the
square root term in~\eqref{eq:56} was chosen. The justification for
doing so is described in Appendix~A.  After substituting this
expression for $\overline{\sigma}_{yx}$ in~\eqref{e-61}, we solve the
equation along with the boundary conditions~\eqref{eq:54} as an
initial value problem by marching from $\xi = 0$ to $\xi = 1$.

Equation~\eqref{e-61} is solved numerically using the \textsc{lsoda}
routine~\citep{petzold83} from \textsc{odepack} in \textsc{netlib}. It
should be noted that the above package estimates $\overline{m}$ at a
small distance $\xi_{1}$ from $\xi = 0$ as
\begin{gather*}
  \overline{m}(\xi_{1}) \approx -NL\nu + \der{\overline{m}}{\xi}(0)\,
  \xi_{1} = -NL\nu,
\end{gather*}
because~\eqref{eq:54} and~\eqref{e-61} imply that
$\der{\overline{m}}{\xi}(0) = 0$. This causes the term under the
square root in~\eqref{eq:56} to be negative. To avoid this
problem~\eqref{e-61} is integrated numerically from $\xi=\xi_{1}$ to
$\xi = 1$, with the initial condition given by
\begin{equation*}
  \overline{m}(\xi_{1}) = -NL\nu + \Frac{1}{2}
  \dertwo{\overline{m}}{\xi}(0)\,\xi_{1}^{2},
\end{equation*}
with $\xi_{1}=10^{-5}$. The use of a smaller value of $\xi_{1}$ does
not significantly affect the results.

Here $\dertwo{\overline{m}}{\xi}(0)$ is calculated by
differentiating~\eqref{e-61} with respect to $\xi$, and
using~\eqref{eq:19} and~\eqref{eq:56}. The resulting indeterminate
expression is evaluated using the L'Hospital's rule to get
\begin{gather}
  \label{eq:20}
  \epsilon\dertwo{\overline{m}}{\xi}(0) = (\mathrm{A} + 1)\nu \mp
  \nu\mysqrtp{\mathrm{A}^{2}-1+
    2(\mathrm{A} + 1)\Frac{N}{L\nu}\dertwo{\overline{m}}{\xi}(0)}.
\end{gather}
This can be rearranged to get a quadratic equation for
$\dertwo{\overline{m}}{\xi}(0)$, and we choose the root that satisfies
\begin{gather*}
  \epsilon\dertwo{\overline{m}}{\xi}(0) \leq (\mathrm{A} + 1)\nu,
\end{gather*}
as the other root is inconsistent with~\eqref{eq:a2}.

Once the stresses are obtained, the velocities are calculated by
integrating the flow rule~\eqref{e-59}) and~\eqref{e-60} from $\xi = 1$ to
$\xi = 0$ using the initial conditions~\eqref{eq:7}. The integration
is started from $\xi = 1$ because
\begin{gather}
  B(\xi)\equiv \Frac{2(1 + \mathrm{A})\,\overline{m}}{\epsilon
    L^{2}(\overline{\sigma}_{yx} +
    \mathrm{A}\overline{\sigma}_{yx})}\label{eq:18}
\end{gather}
becomes unbounded as $\xi\rightarrow0$, and hence the right hand side
of~\eqref{e-60} is indeterminate at $\xi = 0$.

Equation~\eqref{e-60} is therefore integrated from $\xi = 1$ to get
\begin{gather}
  \label{eq:9}
  \overline{\omega}(\xi) = \overline{\omega}_{w}\exp\left(-
  \Int{\xi}{1}{B(\xi')}{\xi'}\right),
\end{gather}
where $\overline{\omega}_{w} =
\omega_{w}(\textfrac{W}{g})^{\textfrac{1}{2}}$ is the dimensionless
angular velocity at the wall.  It is shown in Appendix~B that
$\Int{\xi}{1}{B(\xi')}{\xi'}$ becomes unbounded as $\xi \rightarrow
0$. Hence $\overline{\omega}$ satisfies the boundary condition
$\overline{\omega}(0) = 0$ for all finite values of
$\overline{\omega}_{w}$.

It is also of interest to determine the behaviour of the solutions in
the limit $\epsilon\equiv\textfrac{d_{p}}{W}\rightarrow0$. The issue
here is the scaling of the shear layer thickness as a function of the
channel half-width, $W$. For small $\epsilon$, an asymptotic solution
is constructed using a perturbation technique described in Appendix~C.
The predictions of this solution are discussed in the next section
along with the numerical results.

\subsubsection{Parameter values}
\label{sec:Parameter-values}

The intrinsic density of the particles ($\rho_{p}$) was taken from the
studies of \citet{neddermanandlaohakul80}, \citet{natarajanetal95}
and~\citet{tuzunandnedderman85}. Glass beads were used in all the
experiments. For want of data, the angle of internal friction $\phi$
was taken to be equal to the reported angle of repose.

The parameters $\nu_{\min}$ and $\nu_{\max}$ were chosen to be 0.5 and
0.65, respectively.  The parameters in~\eqref{eq:24} were estimated as
follows.  \citet{jyotsnaandrao97} used the data of~\citet{fickieetal89}
to obtain an expression for the variation of the mean stress at a
critical state ($\sigma_{c}$) as a function of the solids fraction
$\nu$. This expression was used to generate the values of
$\overline{\sigma}_{c}$ for $\nu$ in the range 0.54--0.58, and the
latter were used to estimate $\textfrac{\Lambda}{(\rho_{p} g W)}$, $p$
and~$q$ by the method of nonlinear least squares, using the
Marquardt-Levenberg algorithm~\citep[p.~678]{pressetal92}.  The
parameter values were found to be $\textfrac{\Lambda}{(\rho_{p} g W)}
= 817$, $p = 2.5$, and~$q = 2.2$.

In the experiments of~\citet{neddermanandlaohakul80} and
\citet{natarajanetal95}, a layer of particles was stuck to the walls of
the channel. This will be referred to as a fully rough wall.  When we
compare model predictions with their data, the angle of wall friction
is chosen as $\delta = \tan^{-1} (\sin \phi)$~\citep{kaza82}. For
comparing the predictions with stress measurements
of~\citet{tuzunandnedderman85}, the measured angle of wall friction,
$\delta = 10^{\circ}$, was used.

The value of the parameter $L$, which occurs in the yield
condition~\eqref{eq:8} was estimated to be $10$ by matching predicted
velocity profiles with the data of \citet{neddermanandlaohakul80}~(see
\fig{2}). This value was used in comparisons with all other data. The
parameter $K$ was set to 0.5.

\section{Results}
\label{sec:results}

In this section we compare the predictions of the theory with the data
of~\citet{neddermanandlaohakul80}, \citet{natarajanetal95},
and~\citet{tuzunandnedderman85}.

\subsection{Velocity profiles}
\label{sec:velocity}

\begin{figure}[htbp]
\label{fig:2}
  \begin{center}
    \mypsfragr{y}{{\Large $\textfrac{u}{u(0)}$}}
    \mypsfrag{x}{{\Large $\xi = x/W$}}
    \includegraphics[width=5in]{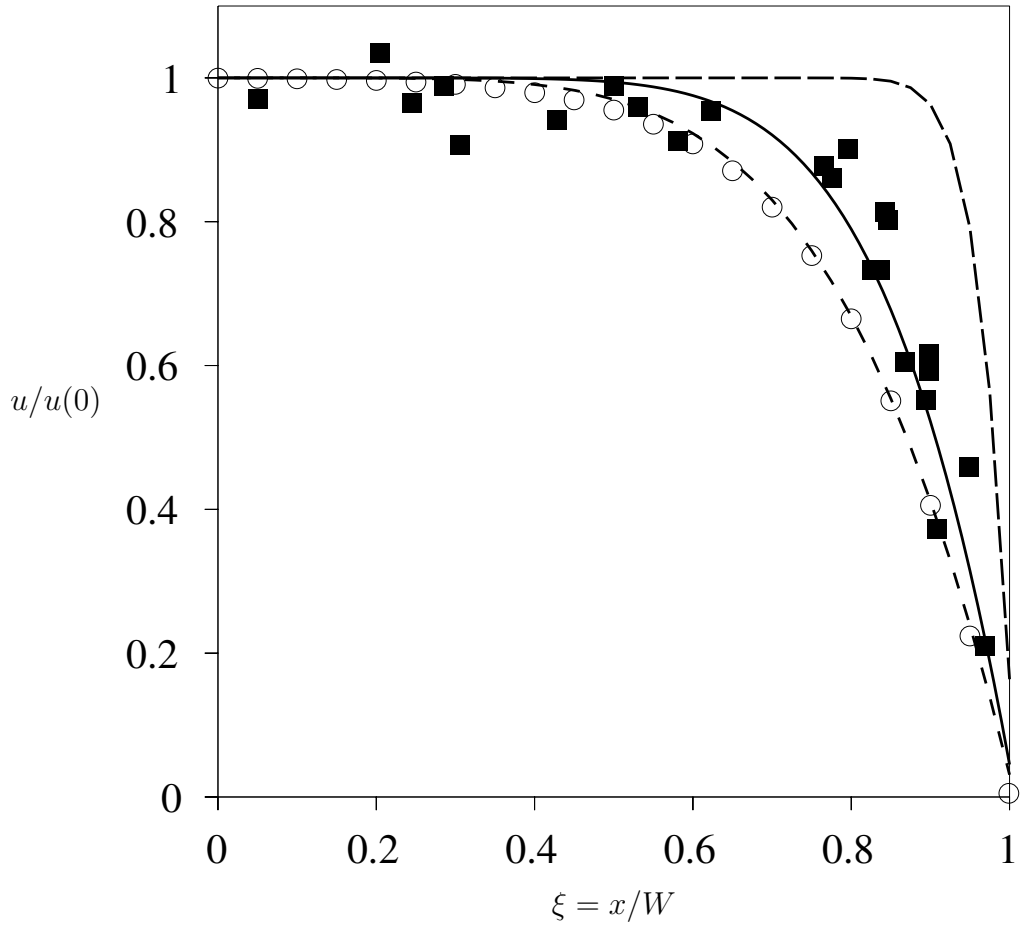}
    \caption{Scaled velocity profiles:
      ($\circ$)~asymptotic solution, ($\BlackBox$)~data of
      \citet{neddermanandlaohakul80} for glass beads. The curves
      represent numerical solutions with $L = 2$~(\linetwo),
      $10$~(\linezero) and $20$~(\lineone).  Parameter values:
      $\mathrm{A} = \textfrac{1}{3}$, $K = 0.5$, $W =
      0.06~\mathrm{m}$, $d_{p} = 0.002~\mathrm{m}$,
      ($\epsilon=\textfrac{1}{30}$), $u(\xi = 0) = 0.2$, $\rho_{p} =
      2940~\mathrm{kg/m^{3}}$, and $\phi = 25^{\circ}$~($\delta =
      22.91^{\circ}$).}
  \end{center}
\end{figure}

With $L = 10$, there is a good match between predicted and measured
linear velocity profiles of~\citet{neddermanandlaohakul80} (\fig{2}).
(Predictions of the theory with $L = 2$ and $L = 20$ are also shown in
\fig{2} for comparison.)  The solids fraction of 0.60 predicted by the
model is in close agreement with the measured average value of 0.61.

While there is no sharply defined plug layer in the model (and in the
experiments), there is a region of low shear rate near the center of
the channel. In order to compare predictions with the data
of~\citet{neddermanandlaohakul80}, the apparent thickness of the
``plug'' layer, $\xi_{p}$ is calculated from
\begin{equation}
  \label{eq:13}
  \Frac{u(\xi_{p})}{u(\xi = 0)} = 0.95.
\end{equation}
Hence the shear layer thickness, scaled by the particle diameter is
$\Delta\equiv(1 - \xi_{p})(\textfrac{W}{d_{p}})$.  The model predicts a
central plug layer and a shear layer adjacent to the wall whose
thickness is about 10.5 particle diameters.

\begin{figure}
\label{fig:3}
  \begin{center}
    \mypsfragr{y}{{\Large $\textfrac{u}{u(0)}$}}
    \mypsfrag{x}{{\Large $\xi = x/W$}}
    \includegraphics[width=5in]{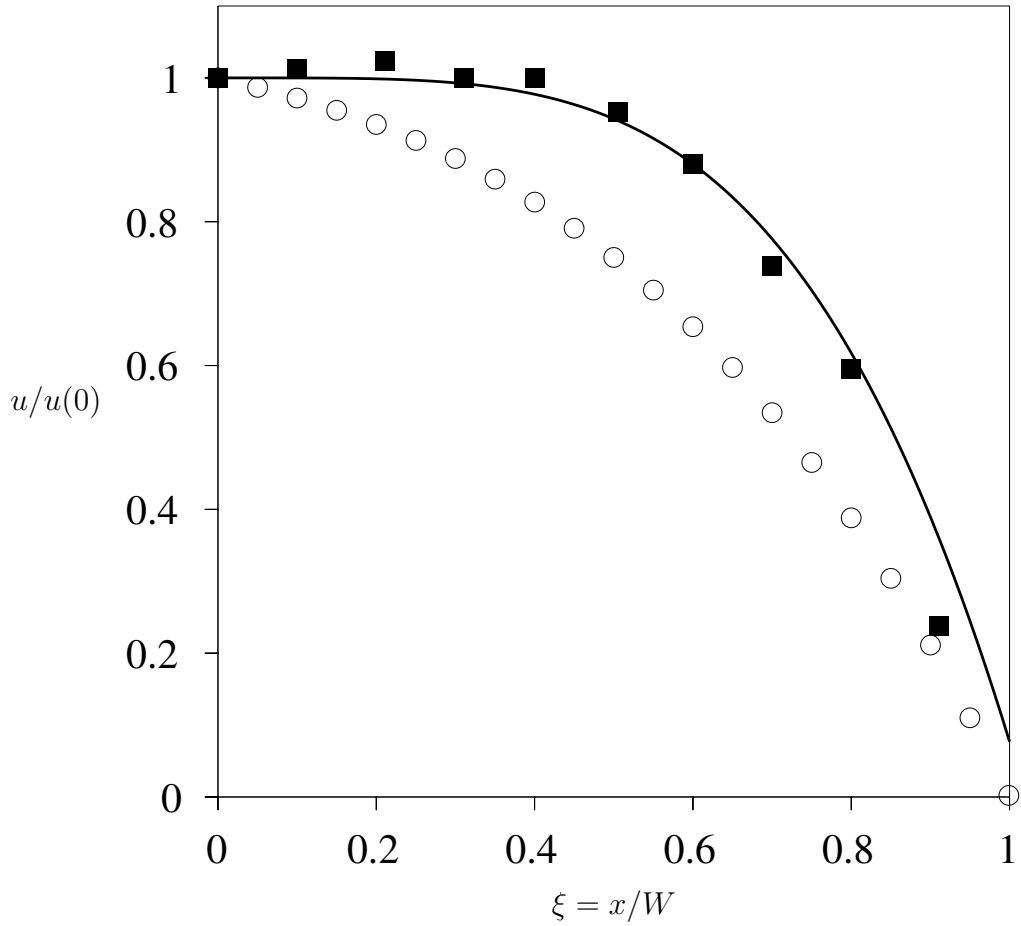}
    \caption{Scaled velocity profiles: (\linezero)~numerical solution,
      ($\circ$)~asymptotic solution ($\BlackBox$)~data of
      \citet{natarajanetal95} for glass beads.  Parameter values: $W =
      0.0255~\mathrm{m}$, $d_{p} = 0.003~\mathrm{m}$, ($\epsilon =
      0.12$), dimensionless mass flow
      rate~$\Int{0}{1}{u\nu}{\xi}=0.15$, $\phi = 28^{\circ}$~($\delta
      = 25.15^{\circ}$), $L = 10$, the rest as in figure~2.}
  \end{center}
\end{figure}

With $L = 10$, the predicted velocity profile also agrees well with
the data of~\citet{natarajanetal95} as shown in~\fig{3}.  This is an
encouraging result because the ratio of the channel width to the
particle diameter differs by a factor of 3.5 for the two sets of data.
For the profile shown in \fig{3}, the solids fraction of 0.59 lies in
the range 0.55--0.67 estimated from the experiments.

\begin{figure}
\label{fig:4}
  \begin{center}
    \mypsfragr{ylabel}{{\Large $\textfrac{u}{u(0)}$}}
    \mypsfrag{xlabel}{{\Large $\xi = x/W$}}
    \includegraphics[width=5in]{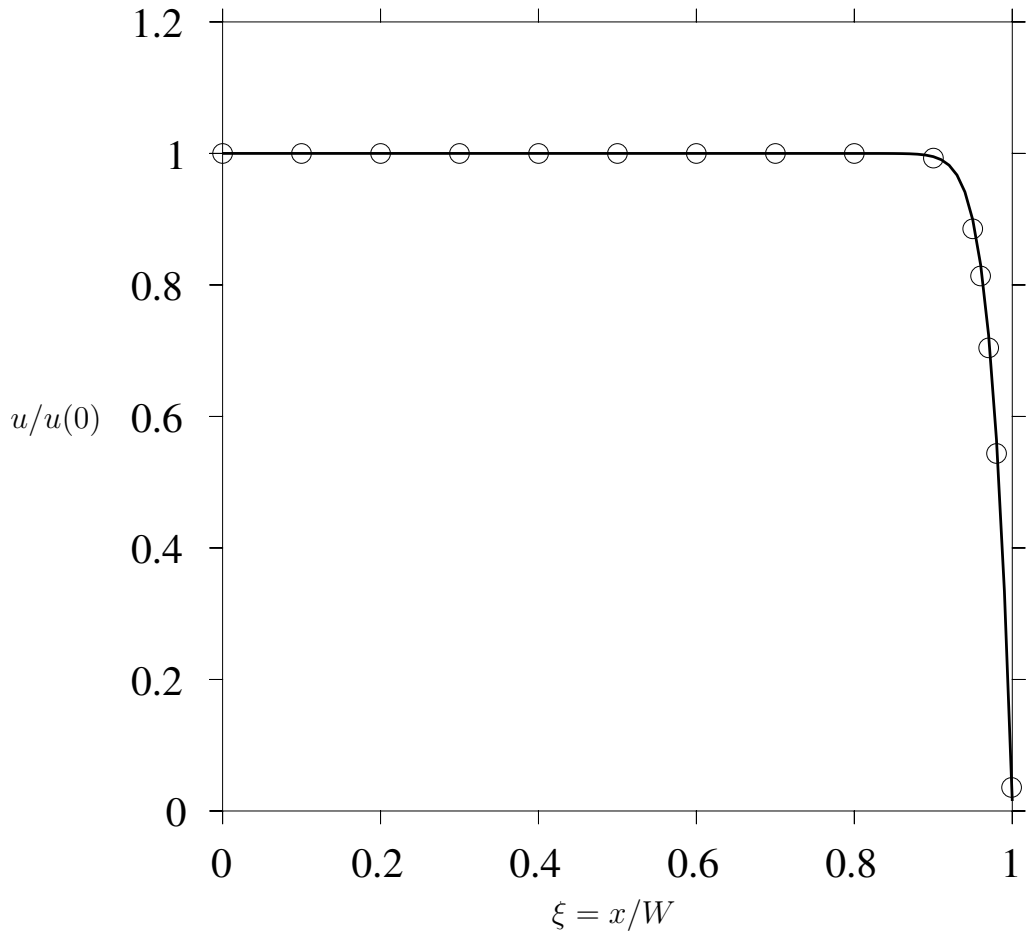}
    \caption{Scaled velocity profiles: (\linezero)~numerical solution,
      ($\circ$)~asymptotic solution. Parameter values: $W =
      1.2~\mathrm{m}$, ($\epsilon=\textfrac{1}{600}$), $L=10$, the
      rest as in figure~2.}
  \end{center}
\end{figure}

The open circles in figures~2 and~3 show the asymptotic velocity
profiles for small $\epsilon$ --- the deviation from the numerical
solution is greater in \fig{3} because $\epsilon$ is larger than that
in \fig{2}. For $\epsilon=\textfrac{1}{600}$, the asymptotic solution
is indistinguishable from the numerical solution, as shown in
\fig{4}.

\begin{figure}
\label{fig:5}
  \begin{center}
    \mypsfrag{x}{{\Large $\xi = x/W$}}
    \psfrag{y}[1][t][1][270]{\shortstack{\Large $\overline{\omega}$,\protect\\
        \Large $\Frac{1}{2}\der{u}{\xi}$}}
    \includegraphics[width=5in]{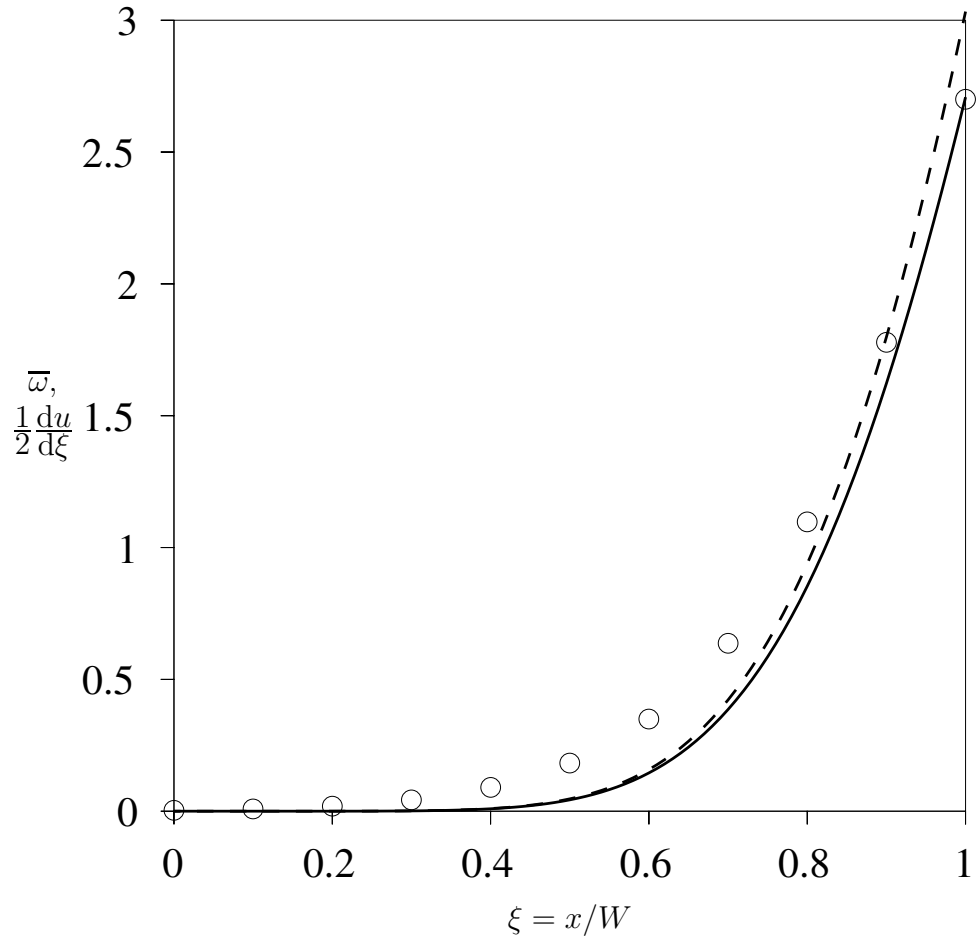}
    \caption{Profiles of the angular velocity
      (\linezero)~$\overline{\omega}$, (\lineone)~half the dimensionless
      vorticity, $\textfracp{1}{2} \textder{u}{\xi}$, and
      the asymptotic solution for $\overline{\omega}$~($\circ$).
      Parameter values: $L=10$, the rest as in figure~2.}
  \end{center}
\end{figure}

\begin{figure}
  \label{fig:6}
  \begin{center}
    \mypsfrag{x}{{\Large $\xi = x/W$}}
    \psfrag{y}[1][t][1][270]{\shortstack{\Large $\overline{\omega}$}}
    \includegraphics[width=5in]{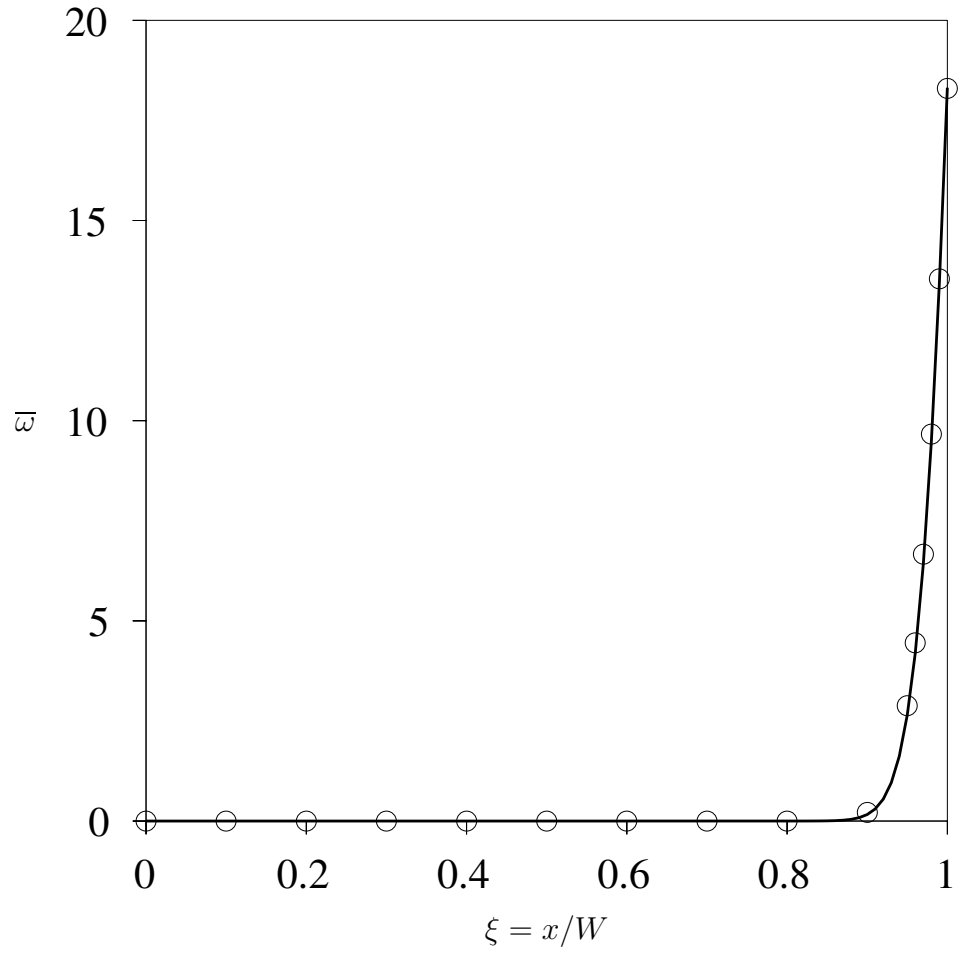}
    \caption{Profiles of the angular velocity:
      (\linezero)~$\overline{\omega}$, and ($\circ$)~the asymptotic
      solution for the angular velocity. Parameter values: $W =
      1.2~\mathrm{m}$, ($\epsilon=\textfrac{1}{600}$), $L = 10$, the rest
      as in figure~2.}
  \end{center}
\end{figure}

The angular velocity ($\overline{\omega}$) profile, shown in \fig{5},
differs slightly from that of half the dimensionless vorticity
$\textfracp{1}{2}\:\textder{u}{\xi}$. As expected, the difference is
more pronounced in the shear layer. (In a classical continuum,
$\textfracp{1}{2}\:\textder{u}{\xi}$ represents the local angular
velocity of an infinitesimal spherical material volume.) The
asymptotic solution deviates significantly from the numerical solution
for $\epsilon=\textfrac{1}{30}$~(\fig{5}), but the two solutions agree
well for $\epsilon=\textfrac{1}{600}$ (\fig{6}).

\subsection{Influence of channel width on the thickness of the shear
  layer}
\label{sec:Infl-chann-width}

For a fixed value of the particle diameter $d_{p}$, the thickness of
the shear layer $\Delta$ increases with the half-width of the
channel~$W$ (solid line in~\fig{7}). This is roughly in accord with
the data of~\citet{neddermanandlaohakul80}, which are represented by
solid symbols in \fig{7}. For each value of the $\textfrac{W}{d_{p}}$,
there are three data points; these correspond to the estimates of
$\Delta$ obtained by fitting three different functional forms to the
measured velocity profile.

For small values of $\epsilon=\textfrac{d_{p}}{W}$, the perturbation
solution~(Appendix~C) shows that
\begin{equation*}
  \Delta \sim \left(\Frac{L^{2}}{2}\right)^{\textfrac{1}{3}}
  \epsilon^{\textfrac{-1}{3}}.
\end{equation*}
Thus the dimensional thickness of the shear layer is proportional to
$(\textfrac{d_{p}}{W})^{\textfrac{-1}{3}}$ when $\textfrac{d_{p}}{W}
\ll 1$, and hence does not attain a constant value in this limit. It
would be interesting to conduct experiments with much larger values of
$\textfrac{W}{d_{p}}$ than in the range shown in \fig{7}. This would
permit a more stringent test of the model predictions.

\subsection{Influence of the parameter $L$ on the thickness of the
  shear layer}

As mentioned earlier, the length scale $L d_{p}$ was chosen to fit the
model predictions to the data of~\citet{neddermanandlaohakul80}. It is
important to know how the predictions vary with changes in this
parameter. Figure~8 shows that the thickness of the shear layer is a
weak function of~$L$. In the limit of small $\epsilon$, the shear
layer thickness varies as $L^{\textfrac{2}{3}}$ (Appendix~C).

\subsection{Influence of the wall-roughness factor $K$ on
  thickness of the shear layer}

The variation of the shear layer thickness with the roughness
parameter $K$ is shown in figure~9.  As mentioned earlier, $K \to
\infty$ corresponds to a very smooth wall; it decreases as the wall
roughness increases.  Figure~9 shows that there is little variation
with $K$ of the shear layer thickness for small $K$, but significant
variation in the range $\approx$ 1--200.  For $K$ greater than 200,
the velocity at the wall is greater than 95\% of the centerline
velocity. Hence, by our definition \eqref{eq:13}, the thickness of the
shear layer is zero.  As shown in Appendix~C, the shear layer
thickness is independent of $K$ in the limit of small $\epsilon$.

\subsection{Stresses}

\subsubsection{Stress profiles}
Figure~10 shows the profiles of the shear stresses
$\overline{\sigma}_{xy}$ and $\overline{\sigma}_{yx}$ for $\epsilon =
\textfrac{1}{30}$ and $\textfrac{1}{600}$.  It is clear that the
difference between $\overline{\sigma}_{xy}$
and~$\overline{\sigma}_{yx}$ increases with $\xi$. Because
$\overline{\sigma}_{yx} > \overline{\sigma}_{xy}$, the couple stress
$\overline{m}$ also increases with $\xi$~(\fig{11}), in accord
with~\eqref{e-61}.

The open symbols in figures~10 and~11 represent the asymptotic solution
for small $\epsilon$. When $\epsilon=\textfrac{1}{600}$, it is clear
that the asymptotic solution is indistinguishable from the exact
solution, and the difference $\overline{\sigma}_{xy} -
\overline{\sigma}_{yx}$ is also very small.

\subsubsection{Wall stresses}

We now compare the predicted wall stresses with the data
of~\citet{tuzunandnedderman85} \mytable{1}. The normal and shear
stresses are over-predicted, but are of the same order of magnitude as
the measured values. As noted by~\citet{mohanetal97}, the dimensions of
the channel used in the experiments are such that the front and the
back faces may support a significant part of the weight of the
material. Hence the shear stress measured at the side wall is expected
to be less than the prediction, which assumes a channel of infinite
depth. It is interesting to note that the estimate
of~\citet{tuzunandnedderman85} for the average solids fraction is 0.63
and the model predicts a value of 0.625.

\begin{table}
  \begin{center}
    \begin{tabular}{lccccc}
      & \multicolumn{1}{c}{$H$} &
      \multicolumn{2}{c}{Normal stress} & \multicolumn{2}{c}{Shear
      stress}\\
      &&Static&Flowing&Static&Flowing\\
      &0.91&2.7--3.6&4.0--5.1&0.30--0.39&0.29--0.35\\
      Experiments&1.07&2.7--2.9&3.8--4.6&0.30&0.41--0.58\\
      &1.22&2.7--3.6&4.6--5.7&0.52--0.65&0.58--0.76\\
      Cosserat model&---&---&7.97&---&1.12\\
      Kinetic model &---&---&2.51&---&1.15
    \end{tabular}
    \caption{Comparison of predicted wall stresses with the data
      of~\citet{tuzunandnedderman85} for glass beads. Here $H$ is the
      depth measured from the top of the channel. Units: $H$--~m,
      stress--~kN/$\mathrm{m}^{2}$. Parameter values: $W =
      0.155~\mathrm{m}$, $d_{p} = 2.29~\mathrm{mm}$, $\rho_{p} =
      1180~\mathrm{kg}/\mathrm{m}^{3}$, $\phi = 30^{\circ}$, $\delta =
      8^{\circ}$.}
    \label{tab:wallstresses}
  \end{center}
\end{table}

\section{Comparison with other models}

\subsection{The classical frictional model}

The classical frictional model predicts a flat velocity profile. This
is consistent with the profile predicted by the Cosserat model in the
limit $d_{p}\rightarrow0$ for a fixed value of $W$. Further, the
Cosserat continuum reduces to the classical continuum in this limit,
because $m\rightarrow0$ and $\sigma_{xy}\rightarrow\sigma_{yx}$.

\subsection{The kinetic and frictional-kinetic models}
\label{sec:Comp-Coss-model}

\begin{figure}
  \label{fig:7}
  \begin{center}
    \mypsfrag{y}{{\Large $(1 - \xi_{p})\textfrac{W}{d_{p}}$}}
    \mypsfrag{x}{{\Large $W/d_{p}$}}
    \includegraphics[width=5in]{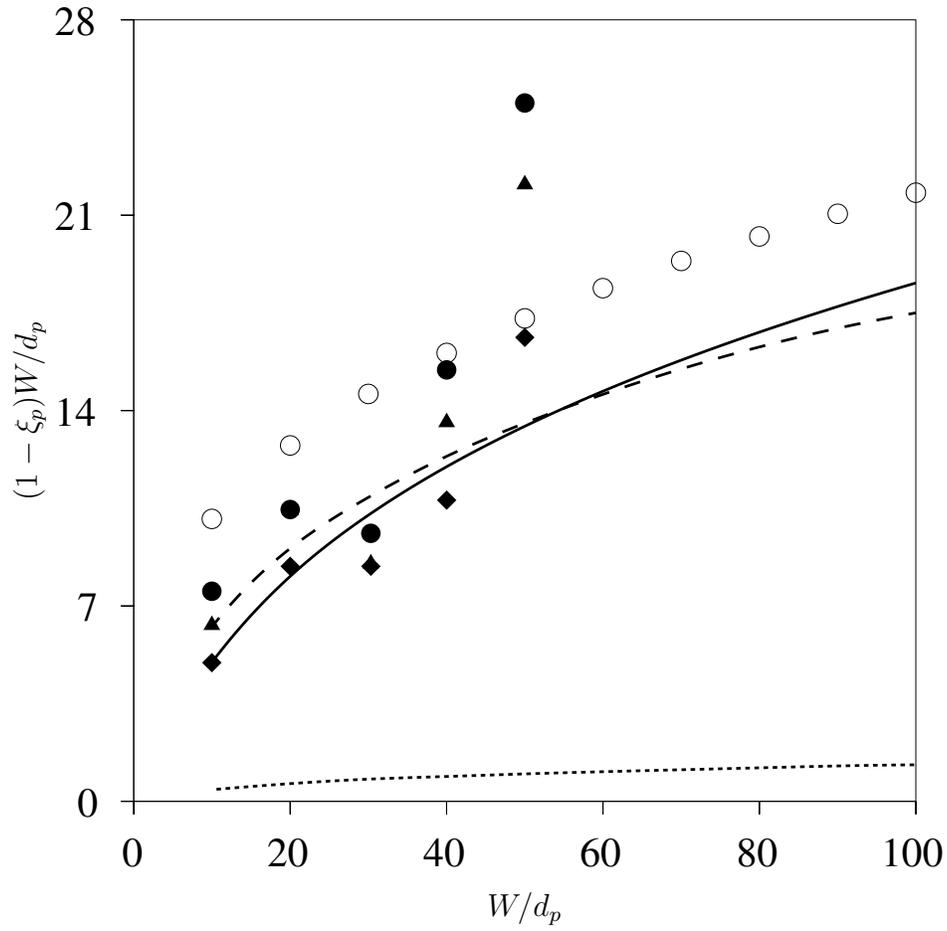}
    \caption{Effect of the channel width $2W$ on the thickness of the
      shear layer (scaled by the particle diameter):
      (\linezero)~numerical solution, ($\circ$)~asymptotic solution,
      (\lineone)~kinetic solution, and (\linedotted)~the
      frictional-kinetic solution. Here the solid symbols represent
      the estimates of \citet{neddermanandlaohakul80}, obtained by
      fitting the data to three different functional forms. Parameter
      values: $L=10$, the rest as in figure~2.}
  \end{center}
\end{figure}

The broken curves in~\fig{7} show the results obtained by using the
(high density) kinetic model and the frictional-kinetic model. For
these models we have used the equations given in~\citet{mohanetal97},
except that the mean stress at critical state ($\sigma_{c}$ in their
paper) is evaluated using~\eqref{eq:24}.

The kinetic model is based on constitutive equations derived by using
the kinetic theory of dense gases~\citep[see, for
example,][]{lunetal84}. Two of the underlying assumptions of this
theory, namely instantaneous binary collisions between particles and
molecular chaos with respect to particle velocities, are expected to
break down at high solids fractions. Therefore it is surprising, and
perhaps fortuitous, that the predicted thickness of the shear layer
is in fair agreement with the data~(\fig{7}) even though the solids
fraction is in the range of 0.64--0.65.

Based on the results shown in~\fig{7}, it is difficult to discriminate
between the Cosserat and the kinetic models. It should be noted that
both these models contain a material length scale in their
constitutive equations. As noted by~\citet{tejchmanandwu94}, this may
be a pre-requisite for a satisfactory description of shear layers.

The frictional-kinetic model is constructed by assuming that the
stress tensor is the sum of the kinetic  stress tensor and the
frictional stress tensor.  This model grossly underestimates the
thickness of the shear layer (see the dotted curve in \fig{7}),
probably because (i)~frictional effects dominate kinetic effects in
the shear layer, and~(ii)~the frictional constitutive equations do not
contain a material length scale.

\subsection{The model of \citet{tejchmanandgudehus93}}
\label{sec:Comparison-with-work}

The work of \citet{tejchmanandgudehus93} appears to be the only other
study which uses a Cosserat model for channel flow.  They use an
elasto-plastic model to examine the batch discharge of material from a
cylindrical bin. The constitutive equations involve the Jaumann stress
rate and the `velocity strain' tensor.  Since they do not present
results for steady fully developed flow, a direct comparison of our
predictions with theirs is not possible.  We are currently attempting
to use their model for the problem at hand, but some issues require
consideration before results can be obtained. For example,
\citet{dienes79} have reported that the Jaumann stress rate furnishes
an unrealistic oscillatory response in simple shear for a hypoelastic model.

\section{Discussion}
\label{sec:Discussion}

Unlike the classical frictional model~\citep{mohanetal97}, the present
Cosserat model predicts velocity profiles which agree well with the
data of~\citet{neddermanandlaohakul80} and~\citet{natarajanetal95}.
Further, the variation of the thickness of the shear layer with the
width of the channel is also captured reasonably well by the model.
The predicted wall stresses are of the same order of magnitude as the
measured values, but there is considerable scope for improvement. In
this context, it may be desirable to account for the finite spacing
between the front and back walls.

Our solution of the model for the limiting case of an infinitely wide
channel~($\epsilon\rightarrow 0$) with fully rough walls indicates
that the shear layer thickness, scaled by the particle diameter, grows
as $\epsilon^{\textfrac{-1}{3}}$, where $\epsilon$ is the ratio of the
particle diameter to the channel width. It would be interesting to
compare this result with experiments conducted for a wide range of
$\epsilon$.

In the present work, and in most applications of the frictional
Cosserat models, \textit{ad hoc} values are prescribed for the
parameters~$a_{1}$, $a_{2}$, and~$L$ in the yield
condition~\eqref{e-10}. Either suitably designed experiments, or a
micro-mechanical treatment, would be valuable in providing estimates
for these parameters. Similarly, it would be desirable to have a
micro-mechanical basis for the kinematic boundary
condition~\eqref{e-63}. An unsatisfactory feature of our model is that
the solids fraction is constant across the channel. This is in
variance with qualitative observations of~\citet{natarajanetal95} that
the density in the shear layer is lower than that in the plug region.
Perhaps the inclusion of elastic or kinetic effects in the model would
correct this feature. In any case, accurate density measurements in
channel flow are lacking, and more investigations in this direction
are needed.

\appendix
\section{Choice of signs in~\eqref{eq:12} and \eqref{eq:56}}
\label{sec:Calculating-limits}

In sections~\ref{sec:problem} and~\ref{sec:solution} it was mentioned
that only the negative root of~\eqref{eq:12} was chosen in our
calculations, and that only one of two possible choices was made in
the sign for the square root in~\eqref{eq:56}. In this appendix we
discuss the justification for doing so.

Rewriting the angular momentum balance~\eqref{e-61} by substituting
for $\overline{\sigma}_{yx}$ from~\eqref{eq:56}, we get
\begin{gather}
  \label{eq:a1}
  \epsilon\der{\overline{m}}{\xi} = (\mathrm{A} + 1)\nu\xi +
  \mysqrt{D}
  \equiv E_{+},\\
  \epsilon\der{\overline{m}}{\xi} = (\mathrm{A} + 1)\nu\xi -
  \mysqrt{D}
  \equiv E_{-},\label{eq:a2}\\
  \intertext{where}
  D = (\mathrm{A}^{2} - 1)\nu^{2}\xi^{2} +
  2(\mathrm{A} + 1)\left(N^{2}\nu^{2} -
    \Frac{\overline{m}^{2}}{L^{2}}\right)\label{eq:a3},
\end{gather}
and $N\equiv\textfrac{\sin\phi}{\tan\delta}$. Equations~\eqref{eq:a1}
and \eqref{eq:a2} have to be integrated subject to the initial
condition
\begin{gather}
  \label{eq:16}
  \overline{m}(0) = \pm N L \nu,
\end{gather}
The qualitative behaviour of the solutions to~\eqref{eq:a1},
\eqref{eq:a2}, and \eqref{eq:16} may be understood by examining the
phase plane of~\eqref{eq:a1} and \eqref{eq:a2}, such as that shown in
\fig{12}.

For $\mathrm{A} < 1 - 2N^{2}$, $D<0$ at $\xi=1$. Hence~\eqref{eq:a1}
and~\eqref{eq:a2} cannot be integrated till the wall ($\xi=1$).

For $\mathrm{A} > -1$ a real valued solution for~\eqref{eq:a1} cannot
be constructed using the initial condition $\overline{m}(0) = N L
\nu$, because $E_{+}\geq 0$ in the region bounded by the curves $D=0$,
and $D<0$ outside this region (see \fig{12}). Suppose that the other
initial condition $\overline{m}(0)=-NL\nu$ is used. For sufficiently
small values of $\epsilon$, the trajectory touches the upper curve
$D=0$ before reaching the channel wall (see, for example, the
dot-dashed line in figure~12).  Hence~\eqref{eq:a1} is discarded.
Using a similar approach, it follows that a suitable solution can be
constructed for~\eqref{eq:a2} only when the initial condition
$\overline{m}(0) = -NL\nu$ is used. A typical trajectory is shown by
the dotted line in \fig{12}.  This choice of roots works only for
small values of $\epsilon$; for the parameters used in \fig{12},
solutions could not be constructed for $\epsilon=0.33$.

For $1-2N^{2}<A<-1$, and small enough $\epsilon$, it is possible to
construct a solution for~\eqref{eq:a1}, subject to the initial
condition $\overline{m}(0) = NL\nu$.

For $A=-1$, it follows from~\eqref{eq:a1} and~\eqref{eq:a2}, and
\eqref{e-61} that $\overline{m} = \mathrm{constant}=-NL\nu$,
$\overline{\sigma}_{xy} = \overline{\sigma}_{yx} = -\nu\xi$, and
the velocity profiles are given by
\begin{align*}
  \overline{\omega} &=
  \overline{\omega}_{w}\,\xi^{\textfrac{2N}{(\epsilon L)}},\\
  u & = u(0) -
  \Frac{2\overline{\omega}_{w}\,\xi^{\textfrac{2N}{(\epsilon L)} + 1}}%
    {\textfrac{2N}{(\epsilon L)} + 1}.
\end{align*}
The other root $\overline{m}(0) = NL\nu$ is discarded because
$\textfrac{\overline{\omega}(0)}{\overline{\omega}_{w}}\rightarrow
\infty$ as $\xi\rightarrow0$.

\section{Integration of~\eqref{e-60}}
\label{sec:Integration}

In \S~\ref{sec:Method-solution} it was noted that
$\overline{\omega}(0)=0$ because
$\lim_{\xi\rightarrow0}\Int{\xi}{1}{B(\xi')}{\xi'} = \infty$, where
$B(\xi)$ is defined by~\eqref{eq:18}. This is shown below.

Near $\xi = 0$, the leading order behaviour of the stresses can be
represented as
\begin{gather*}
  \overline{\sigma}_{xy} = -\nu\xi,\\
  \overline{\sigma}_{yx} =
  \overline{\sigma}_{yx,0}'\,\xi
  +\OrderOf{\xi^{2}},\\
  \overline{m} = \overline{m}(0) + \OrderOf{\xi^{2}},
  \intertext{where}
  \overline{\sigma}_{yx,0}'\equiv
  \left.\der{\overline{\sigma}_{yx}}{\xi}\right\vert_{\xi=0}.
\end{gather*}

The integral can now be written as
\begin{align}
  \label{eq:21}
  \lim_{\xi\rightarrow0}\Int{\xi}{1}{B(\xi')}{\xi'} &=
  \lim_{\xi\rightarrow0}
  \Int{\xi}{\alpha}{\Frac{2(\mathrm{A} + 1)\,\overline{m}(0)}%
    {(\overline{\sigma}'_{yx,0} -
      \mathrm{A}\nu)\xi'}}{\xi'}
  + \Int{\alpha}{1}{B(\xi')}{\xi'},\\
  &=\lim_{\xi\rightarrow0}\Frac{2(\mathrm{A} + 1)\,\overline{m}(0)}%
  {(\overline{\sigma}'_{yx,0} -
    \mathrm{A}\nu)}\ln\left(\Frac{\alpha}{\xi}\right) +
  \Int{\alpha}{1}{B(\xi')}{\xi'},\label{eq:25}
\end{align}
where $\alpha$ is a small positive number.  When $\mathrm{A} > -1$,
and $\overline{m}(0)$ and $\overline{\sigma}_{yx}$ are evaluated as
discussed in Appendix~A, it follows from~\eqref{e-61}
and~\eqref{eq:20} that the factor multiplying the logarithm
in~\eqref{eq:25} is positive, and hence the expression
in~\eqref{eq:25} becomes unbounded.

\section{Asymptotic solution for the Cosserat model}
\label{sec:Asympt-solut-Coss}

Here we derive an asymptotic solution for the case of a fully rough
wall (\ie, one for which $\tan\delta=\sin\phi$, or $N=1$), in the
limit $\epsilon\rightarrow0$.

Using~\eqref{eq:7}, \eqref{eq:54}, and \eqref{eq:56},~\eqref{e-61} may
be rewritten as
\begin{equation}
  \label{eq:14}
  \epsilon \der{\overline{m}}{\xi} -
  (\mathrm{A} + 1)\nu\xi + \mysqrtp{(\mathrm{A}^{2} -
  1)\:(\nu\xi)^{2} +
  2(\mathrm{A} + 1)\left(\nu^{2} -
  \Frac{\overline{m}^{2}}{L^{2}}\right)} = 0.
\end{equation}
We now seek a solution for $\overline{m}$ of the form
\begin{equation}
  \label{eq:17}
  \overline{m} = m_{0} + \epsilon m_{1} + \epsilon^{2} m_{2} +
  \ldots.
\end{equation}
Substituting this in~\eqref{eq:14}, expanding the term under the
square root for small $\epsilon$, and collecting terms of \OrderOf{1}
and \OrderOf{$\epsilon$} we get
\begin{equation}
  \label{eq:15}
  m_{0} = -L\nu\mysqrtp{1 - \xi^{2}}; \quad m_{1} =
  -\Frac{L^{2}\nu^{2}\xi^{2}}{2(1 - \xi^{2})}.
\end{equation}
Equation~\eqref{eq:15} shows that the solution for $\overline{m}$ is
not uniformly valid as $\Modulus{\textfrac{\epsilon m_{1}}{m_{0}}}
\sim \OrderOf{1}$ when $\xi\approx 1
-\textfracp{1}{2}\epsilon^{\textfrac{2}{3}}$.

To get an uniformly valid first approximation for $\overline{m}$, we
proceed as suggested by \citet[p.~104]{dyke64}. Introducing new
variables
\begin{gather*}
  \inner{\xi}\equiv(1 - \xi)\,\epsilon^{\textfrac{-2}{3}};\quad
  \inner{m}\equiv\overline{m}\,\epsilon^{\textfrac{-1}{3}},
\end{gather*}
such that they are \OrderOf{1} in the inner region, we seek a solution
of the form
\begin{equation}
  \label{eq:3}
  \inner{m} = \inner{m}_{0} + f(\epsilon)\,\inner{m}_{1} + \ldots.
\end{equation}
Substituting \eqref{eq:3} in~\eqref{eq:14}, expanding the term under
the square root for small $\epsilon$, and collecting terms of
\OrderOf{1}, we get
\begin{equation}
  \label{eq:27}
  \der{\inner{m}_{0}}{\inner{\xi}} = 2 \nu\inner{\xi} -
  \Frac{{\inner{m}_{0}}^{2}}{\nu L^{2}}.
\end{equation}
Equation~\eqref{eq:27} is a Riccati
equation~\citep[p.~20]{benderandorszag84}, which can be converted to a
second-order linear differential equation by using the transformation
\begin{equation}
  \label{eq:28}
  \inner{m}_{0}(\inner{\xi}) = \Frac{\nu L^{2}
  \textder{\Inner{m}_{0}}{\inner{\xi}}}{\Inner{m}(\inner{\xi})}
\end{equation}
to get
\begin{equation}
  \label{eq:36}
  \dertwo{\Inner{m}}{\inner{\xi}} = \Frac{2 \inner{\xi}}{L^{2}}
  \Inner{m}.
\end{equation}
Using the transformation
$\xibar\equiv(\textfrac{2}{L^{2}})^{\textfrac{1}{3}}\:\inner{\xi}$,
\eqref{eq:36} reduces to the Airy equation
\begin{equation}
  \label{eq:37}
  \dertwo{\Inner{m}_{0}}{\xibar} = \xibar \Inner{m}_{0},
\end{equation}
and its general solution is given by~\citep[p.~100]{benderandorszag84}
\begin{equation}
  \label{eq:38}
  \Inner{m}_{0} = C_{1}\airyai\:(\xibar) + C_{2}\airybi\:(\xibar).
\end{equation}
Here $\airyai$ and $\airybi$ are the linearly independent Airy
functions, and $C_{1}$ and $C_{2}$ are integration constants. Hence
\begin{equation}
  \label{eq:39}
  \inner{m}_{0} = \left(\Frac{2}{L^{2}}\right)^{\textfrac{1}{3}}
  \Frac{\nu L^{2}}{C_{1}\airyai(\xibar) +
  C_{2}\airybi(\xibar)}\: \der{}{\xibar}\left(C_{1}\airyai(\xibar) +
  C_{2}\airybi(\xibar)\right).
\end{equation}
To determine $C_{1}$ and $C_{2}$, we follow the procedure discussed
in~\citet[p.~105]{dyke64}. The outer solution~\eqref{eq:15} is
rewritten in terms of the inner variable and expanded for small
$\epsilon$ to get the leading order inner expansion of the outer
solution
\begin{gather*}
  -L\nu\epsilon^{\textfrac{1}{3}}\:\mysqrtp{2 \inner{\xi}}.
\end{gather*}
Similarly, the inner solution should be rewritten in terms of the
outer variables and expanded for small $\epsilon$. This is exactly
equivalent to expanding~\eqref{eq:39} in the limit $\xibar \rightarrow
\infty$. The leading order outer expansion of the inner solution
is
\begin{align*}
  -L\nu\epsilon^{\textfrac{1}{3}}\:\mysqrtp{2 \inner{\xi}}
  &\quad\text{if}\quad C_{2} = 0;\\
  L\nu\epsilon^{\textfrac{1}{3}}\:\mysqrtp{2 \inner{\xi}}
  &\quad\text{if}\quad C_{1} = 0,
\end{align*}
where the asymptotic expansions of the Airy
function~\citep[p.~100]{benderandorszag84} have been used. Thus the
inner and outer expansions have the same functional behaviour in the
``overlap'' region provided $C_{2} = 0$; hence
\begin{equation}
  \label{eq:4}
  \Inner{m}_{0} = C_{1}\airyai\:(\xibar).
\end{equation}
Following the procedure described in~\citet[p.~94]{dyke64}, the
leading order composite (additive) solution is given by
\begin{equation}
  \label{eq:44}
  \overline{m} = -L\nu\mysqrtp{1 - \xi^{2}} +
  \left(\Frac{2\epsilon}{L^{2}}\right)^{\textfrac{1}{3}}
  \Frac{1}{\airyai(\xibar)}\der{\airyai}{\xibar} +
  \nu L\mysqrtp{2(1 - \xi)}.
\end{equation}
Similarly, the composite solution for the other variables is given by
\begin{align}
  \overline{\sigma}_{yx} &= -2\nu + \nu\xi +
  \left(\Frac{2\epsilon^{2}}{L^{2}}\right)^{\textfrac{1}{3}}
  \Frac{1}{L^{2}\nu\airyai(\xibar)}\der{\airyai}{\xibar},\label{eq:29}\\
  \overline{\omega} &= \overline{\omega}_{w}{\left(
      \Frac{\airyai{(\xibar)}}{\airyai(0)}\right)}^{2},\label{eq:30}\\
  u &= 2\overline{\omega}_{w}
  {\left(\Frac{2\epsilon^{2}}{L^{2}}\right)}^{\textfrac{1}{3}}
  I(\xibar),\label{eq:32}\\
  \intertext{where} I(\xibar) &= \Int{0}{\xibar}
  {{\left(\Frac{\airyai(z)}{\airyai(0)}\right)}^{2}}{z}.\notag
\end{align}
The parameter $\overline{\omega}_{w} = \overline{\omega}(\xi = 1) =
\overline{\omega}(\xibar = 0)$ is determined as follows. Using~\eqref{eq:32}
and the measured centerline velocity $u_{e}(\xi = 0)$, we get
\begin{gather}
  \label{eq:10}
  \overline{\omega}_{w} =
  \Frac{u_{e}(\xi = 0)}%
  {2{\left(\Frac{2\epsilon^{2}}{L^{2}}\right)}^{\textfrac{1}{3}}I(\xibar_{0})},
\end{gather}
where $\xibar_{0} =
(\textfrac{2}{L^{2}})^{\textfrac{1}{3}}\epsilon^{\textfrac{-2}{3}}$.

Using~\eqref{eq:13}, \eqref{eq:32} and \eqref{eq:10}, the thickness
$\xi_{p}$ of the plug is given by
\begin{gather*}
  \Frac{u(\xi_{p})}{u(\xi=0)} = \Frac{I(\xibar_{p})}{I(\xibar_{0})} =
  0.95,
\end{gather*}
where $\xibar_{p}=(1-\xi_{p})\left(\textfrac{2}{L^{2}}\right)^{\textfrac{1}{3}}
\epsilon^{\textfrac{-2}{3}}$.

In the limit $\epsilon\rightarrow0$,
$\overline{\xi}_{0}\rightarrow\infty$ and $I(\overline{\xi}_{0})$
tends to a constant. Hence, for small $\epsilon$, $\overline{\xi}_{p}$
is approximately independent of $\epsilon$, and
$\lim_{\epsilon\rightarrow0}\xibar_{p} = 1.275$. The dimensionless
shear layer thickness is therefore given by
\begin{gather*}
  1 - \xi_{p} =
  \left(\Frac{L^{2}}{2}\right)^{\textfrac{1}{3}}
  \epsilon^{\textfrac{2}{3}}\,\xibar_{p},
\end{gather*}
and the shear layer thickness expressed in terms of particle diameter is
\begin{gather*}
  \Delta\equiv(1 -\xi_{p})\Frac{W}{d_{p}} =
  \left(\Frac{L^{2}}{2}\right)^{\textfrac{1}{3}}
  \epsilon^{\textfrac{-1}{3}}\,\xibar_{p}.
\end{gather*}
The above results are valid for the case of a fully rough wall
($N=\textfrac{\sin\phi}{\tan\delta}=1$). For smoother walls ($N>1$),
the outer solution is uniformly valid in the limit
$\epsilon\rightarrow0$.

\begin{figure}
  \label{fig:8}
  \begin{center}
    \mypsfrag{x}{{\Large $L$}}
    \psfrag{y}[bl]{{\Large $(1 - \xi_{p})\textfrac{W}{d_{p}}$}}
    \includegraphics[width=5in]{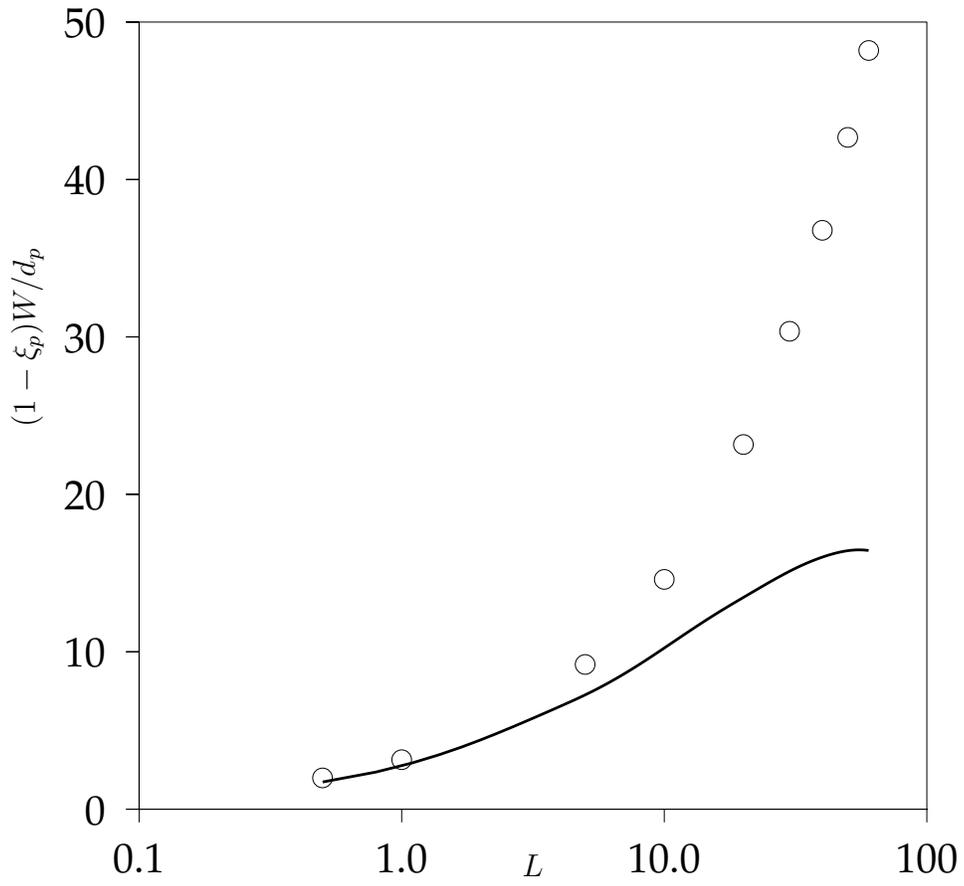}
    \caption{Influence of $L$ on the predicted
      thickness of the shear layer: (\linezero)~numerical solution,
      ($\circ$)~asymptotic solution.  Parameter values as in figure~2.}
  \end{center}
\end{figure}

\begin{figure}
  \label{fig:new}
  \begin{center}
    \mypsfrag{x}{{\Large $K$}}
    \mypsfrag{y}{{\Large $(1 - \xi_{p})\textfrac{W}{d_{p}}$}}
    \includegraphics[width=5in]{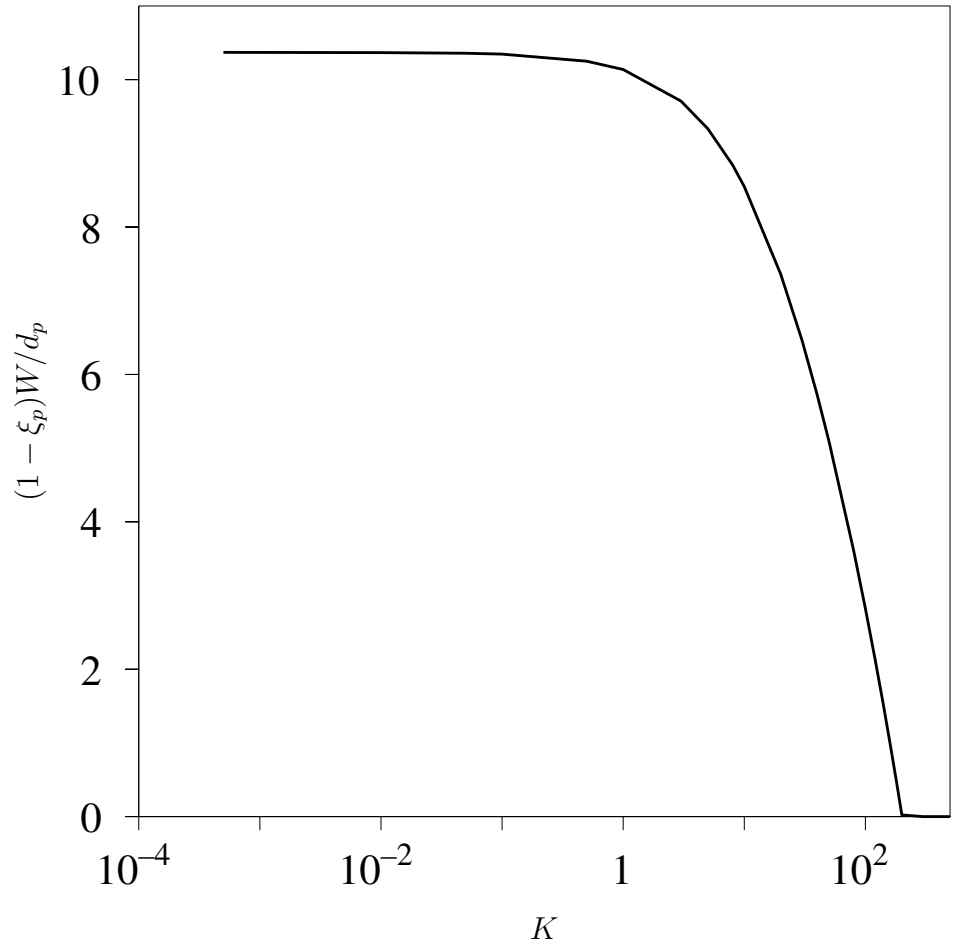}
    \caption{Influence of $K$ on the predicted
      thickness of the shear layer. Parameter values as in figure~2.}
  \end{center}
\end{figure}

\begin{figure}
  \label{fig:9}
  \begin{center}
    \mypsfrag{x}{{\Large $\xi = \Frac{x}{W}$}}
    \psfrag{y}[Tr][1][1][270]{\shortstack{\Large $\overline{\sigma}_{xy}$,\\
        \Large $\overline{\sigma}_{yx}$}}
    \includegraphics[width=5in]{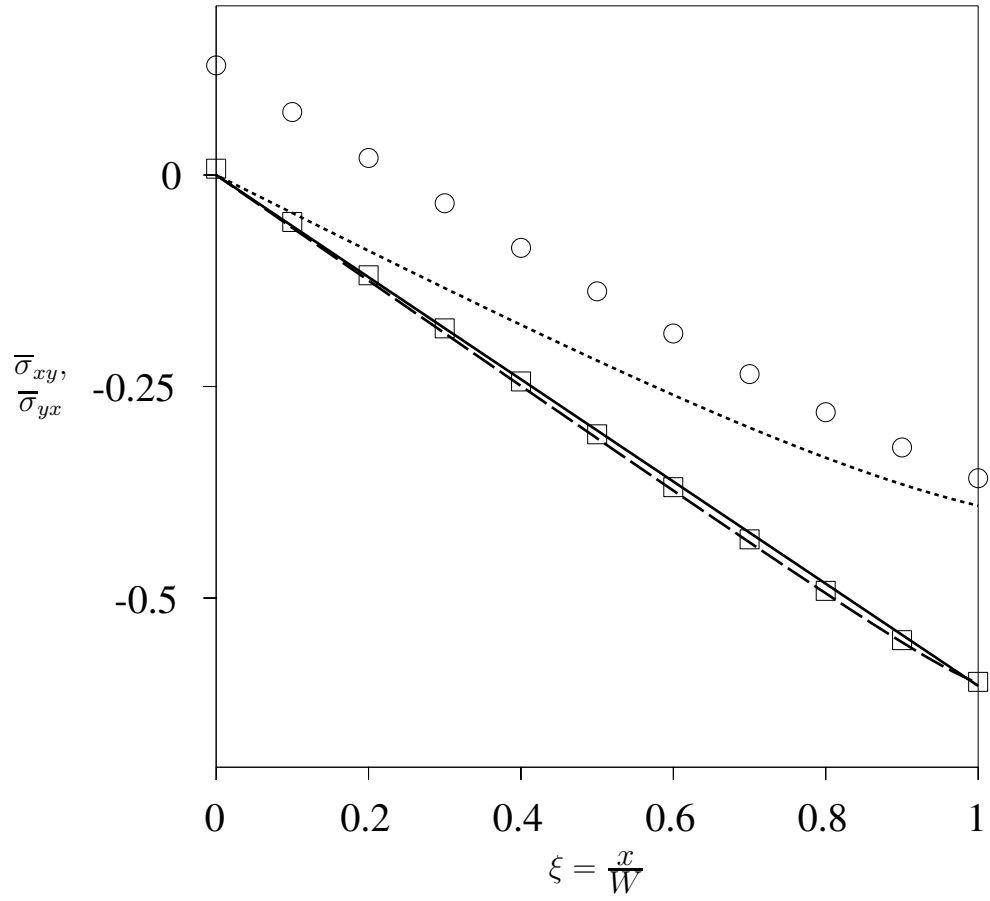}
    \caption{Profiles of the shear stresses:
      (\linezero)~$\overline{\sigma}_{xy}$,
      $\overline{\sigma}_{yx}$~(\linedotted,
      $\epsilon=\textfrac{1}{30}$; \linetwo,
      $\epsilon=\textfrac{1}{600}$). The symbols represent the
      asymptotic solutions for $\overline{\sigma}_{yx}$ ($\circ$,
      $\epsilon=\textfrac{1}{30}$; $\Box$,
      $\epsilon=\textfrac{1}{600}$).  Parameter values: $L=10$, the rest as in
      figure~2.}
  \end{center}
\end{figure}

\begin{figure}
\label{fig:10}
  \begin{center}
    \mypsfrag{x}{{\Large $\xi = \Frac{x}{W}$}}
    \mypsfragr{y}{{\Large $\overline{m}$}}
    \includegraphics[width=5in]{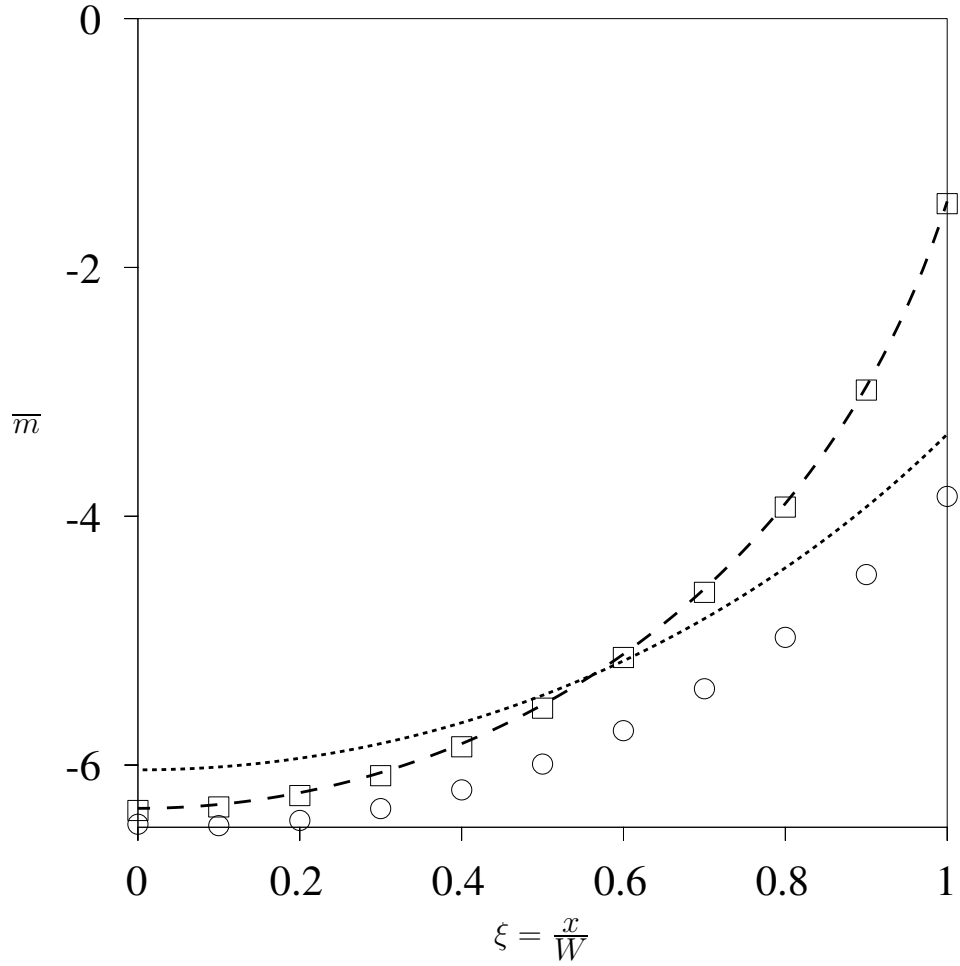}
    \caption{Profiles of the couple stress
      $\overline{m}$~(\linedotted,~$\epsilon=\textfrac{1}{30}$;
      \lineone,~$\epsilon=\textfrac{1}{600}$). The symbols represent
      asymptotic solutions~($\circ$, $\epsilon=\textfrac{1}{30}$;
      $\Box$, $\epsilon=\textfrac{1}{600}$). Parameter values as in
      figure~2.}
  \end{center}
\end{figure}

\begin{figure}
  \label{fig:11}
  \begin{center}
    \mypsfrag{x}{{\Large $\xi$}}
    \mypsfrag{y}{{\Large $\overline{m}$}}
    \mypsfrag{d1}{{\Large $E_{+} > 0$, $E_{-} < 0$}}
    \mypsfrag{d2}{{\Large $D<0$}}
    \mypsfrag{d3}{{\Large $D=0$}}
    \mypsfrag{e0}{{\Large $E_{-} = 0$}}
    \mypsfrag{e1}{{\Large $E_{-} > 0$}}
    \includegraphics[width=5in]{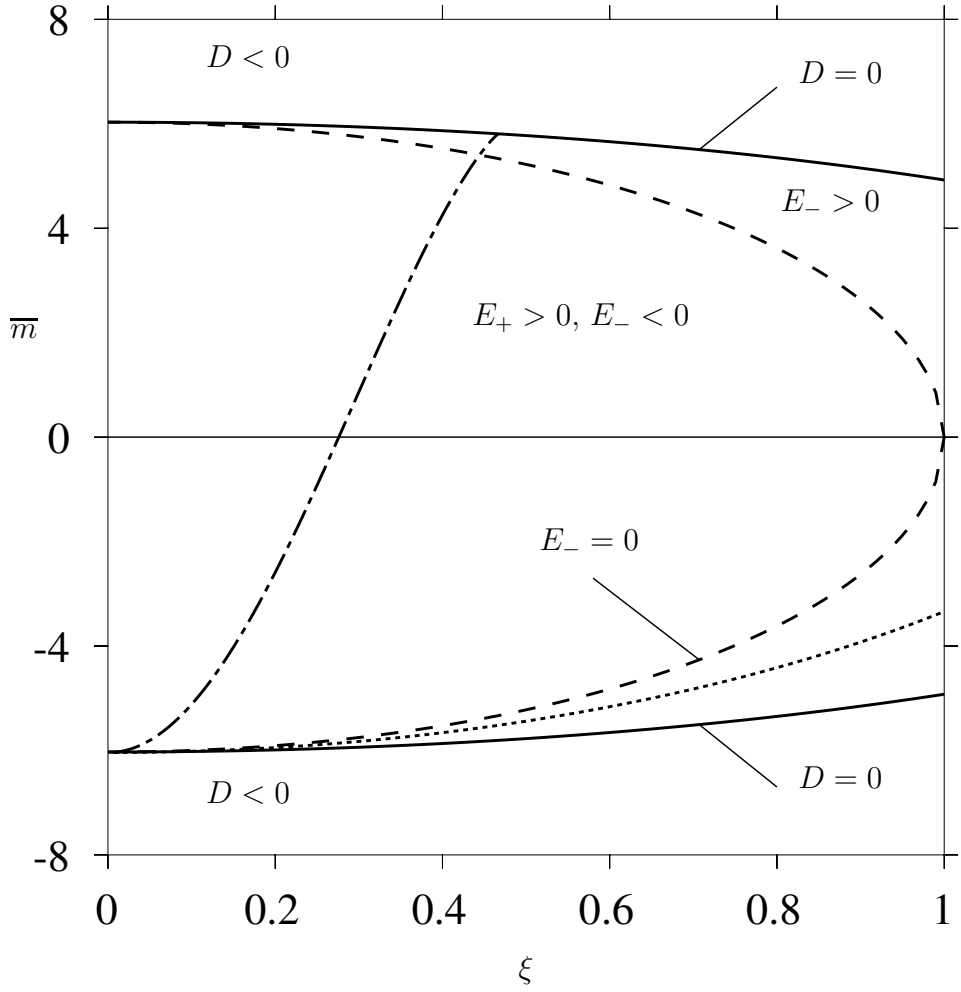}
    \caption{Phase plane of the angular momentum balance
      equations~(A1) and (A2):
      \linedotdash,
      trajectory corresponding to (A1); \linedotted, trajectory
      corresponding to (A2). In both cases, the initial condition
      is $\overline{m}(0) = -NL\nu$. Parameter values: $N = 1$,
      $L=10$, ($\nu=0.603$),
      $\mathrm{A}=\textfrac{1}{3}$, $\epsilon=\textfrac{1}{30}$.}
  \end{center}
\end{figure}


\end{document}